\documentclass[aps,pre,showkeys,manuscript]{revtex4}
\usepackage[centertags]{amsmath}
\usepackage{amssymb}
\usepackage{amsthm}
\usepackage{graphicx,subfigure}
\usepackage{epsfig}
\usepackage{multirow}
\usepackage{color}
\usepackage[utf8]{inputenc}
\usepackage[english]{babel}
\usepackage{cancel}
\usepackage[normalem]{ulem}

\begin{document}
\title{Magnetic field induced asymmetric splitting of the output signal}

\author{L. R. Rahul Biswas, Joydip Das and Bidhan Chandra Bag
{\footnote{Author for correspondence, e-mail:bidhanchandra.bag@visva-bharati.ac.in}}}

\affiliation{Department of Chemistry, Visva-Bharati, Santiniketan 731 235, India}

\begin{abstract}
In this paper we have investigated the dynamics of a damped harmonic oscillator in the presence of an electromagnetic field. The transients for the two dimensional harmonic oscillator imply about the modulation of the frequency of the oscillator by the velocity dependent non conservative force from an applied magnetic field. Except a special condition, the motion is in general quasi periodic nature even in the absence of damping. Another interesting finding is that the magnetic field may induce an asymmetric splitting of the spectrum of the output signal with two peaks in the case of a driven damped two dimensional harmonic oscillator. One more additional peak may appear for the three dimensional case. In some cases the spectrum may have similarity with the Normal  Zeeman Effect. At the same time one may observe to appear the anti resonance phenomenon even for the driven damped cyclotron motion where the system with the purely non conservative force fields is driven by an electric field. Finally, our calculation exhibits how the magnetic field can modulate the phase difference (between input and output signals) and the efficiency like quantity of the energy storing process. Thus the present study might be applicable in the areas related to the refractive index, the barrier crossing dynamics and autonomous stochastic resonance, respectively.
\end{abstract}


\maketitle

\section{Introduction}

Although the study on the dynamics in the presence of an electromagnetic field is an old issue but due to the following facts it may be still an important issue in the recent technology. The investigation on the ion conducting electrolytic materials is a key area in physics and chemistry \cite{scros, gray, wright, bruce, armand, angel, lilley}. The materials have potential applications in a diverse range of all-solid-state devices, such as rechargeable lithium batteries, flexible electrochromic displays and smart windows \cite{scros}. The properties of the electrolytes are tuned by varying chemical composition to a large extent and hence are adapted to specific needs \cite{angel, lilley}. High ionic conductivity is needed for optimizing the glassy electrolytes in various applications. Then it would be very interesting if one can tune the ionic conductivity according to specific need by a physical method. In this context, very recent studies \cite{katsuki, pere, telang, vdo, bag3, bag4, bag5, bag6, physa} show that the conductivity of an electrolytic material can be tuned by an applied magnetic field. To tune the conductivity of ions in the solid electrolytes the combination of both magnetic field and time-dependent electric field may be an important choice. Then one may be interested to know the basic dynamics as well as energetics of an analytically solvable model like driven damped harmonic oscillator in the presence of a magnetic field. The driven damped harmonic oscillator is a well studied text book material \cite{feyn, symon}. At the same time, the dynamics of a particle in the presence of both constant magnetic and electric fields is also a well studied issue in the text book \cite{symon}. This study has been extended in different contexts\cite{extnmf,extnmf1}.
But to the best of our knowledge, the study on the dynamics of periodically driven damped harmonic oscillator in the presence of a magnetic field was not addressed. It does not mean that this issue is not an important one. In other words, the
model study may be a very relevant one in the context of barrier crossing dynamics as mentioned above. At the same time it may also be an important one to explain the refractive     
of a dielectric material in the presence of a magnetic field\cite{refi}. 
Thus our objective is to explore distinguishable feature (if any) of this dynamics including the energetics and the resonance at the steady state. Then we start with the transients of the two dimensional damped harmonic oscillator (having same frequency along both $x$ and $y$-directions) in the presence of a magnetic field. $x(t)$ and $y(t)$ are superposition of two periodic terms in the absence of damping. Thus the motion may be quasi periodic in nature. Then we determine the condition for simple periodic motion. At the same time we determine how the frequencies of the periodic terms in the damped oscillation may depend on the damping strength. These calculations corroborate to the resonance conditions which are determined based on the steady state dynamics for the driven system. Here we find that the magnetic field induces an asymmetric splitting of the spectrum of the output signal with the finite values of the amplitude at the resonance conditions. But the values of the amplitudes at resonance condition become infinite in the absence of damping. It is to be noted here that for this case, the phase shift between the input and output signals at the resonance condition is similar as that of the driven damped harmonic oscillator. It proofs indirectly that the finite value of the amplitude at the resonance condition in the presence of damping is solely due to the dissipation of energy. In other words, the phase shift has no significant role in this context. Major points like these have been included in the conclusion section. 

Before leaving this section we would mention that in recent past the magnetic field has been considered in different contexts such as barrier crossing dynamics \cite{katsuki, pere, telang, vdo, bag3, bag4, bag5, bag6, physa}, non Markovian dynamics of a Brwonian particle in the presence of a magnetic field \cite{bag4, bag6, physa, nmar, nmar1, nmar4, jcp, preha}, stochastic thermodynamics \cite{stoct}, nonlinear dynamics \cite{nonlin} and others \cite{indmag, fabio, jayn, gelf}. The present study may be relevant in some of these areas. In the conclusion section we have mentioned about the possible applications of the present investigation. 

The outlay of the paper is as follows. In Sec.II, we have presented the dynamics of a two dimensional harmonic oscillator in the presence of an electromagnetic field. The steady state dynamics of a driven damped three dimensional harmonic oscillator in the presence of a magnetic field has been addressed in the next section. The paper is concluded in Sec.IV.

\section{Dynamics of a two dimensional harmonic oscillator in the presence of an electromagnetic field}

\subsection{Transients: Implication of the resonance condition}

The dynamics of a oscillating particle (with angular frequency $\omega$ and mass $m$) in the presence of  a magnetic field ${\bf B} = (0,0,B_z)$ and the frictional force can be described by

\begin{equation}\label{eq1}
m \dot{u}_x = - m \omega^2 x + m \Omega u_y - \gamma u_x   \;\;\;,
\end{equation}

and

\begin{equation}\label{eq2}
m \dot{u}_y = - m \omega^2 y - m \Omega u_x - \gamma u_y   \;\;\;.
\end{equation}

\noindent
Here $u_x$ and $u_y$ correspond to the components of the velocity for the motion in the $x$-$y$ plane. The magnetic force from the given field is confined in this plane as implied in the above equations of motion. Here the parameter, $\Omega = \frac{q B_z}{m}$ is the cyclotron frequency for the rotational motion of the particle (with charge, $q$) which is driven by only the magnetic force. The remaining parameter, $\gamma$ in Eqs. (\ref{eq1}-\ref{eq2}) measures the damping strength. However, the above coupled equations of motion can be solved using the transformation, $\xi = x + iy$ \cite{landau}. Then we have

\begin{equation}\label{eq3}
\ddot{\xi} = - \omega^2 \xi - \beta \dot{\xi}   \;\;\;,
\end{equation}

\noindent
where $\beta = \gamma + i \Omega$. This leads to have the solution of the above equation as

\begin{equation}\label{eq4}
\xi(t) = \xi(0) e^{- \beta/2 t} \cos\left(\sqrt{\omega^2 - \beta^2/4} t\right)   \;\;\;.
\end{equation}

\noindent
Here $\xi(0)$ can be identified as, $x(0) + iy(0)$. However, the argument of $\cos(\sqrt{\omega^2 - \beta^2/4} t)$ (which may be represented by $\theta$) restricts us to proceed further analytically. It can be expressed as

\begin{equation}\label{eq4a}
\theta = \sqrt{A + iB} t   \;\;\;.
\end{equation}

\noindent
Here we have used $A = \omega^2 - \gamma^2/4 + \Omega^2/4$ and $B = - \frac{\gamma \Omega}{2}$. If $\gamma$ as well as $B$ is zero then decomposing the above solution one may read time dependence of $x(t)$ and $y(t)$ as

\begin{eqnarray}\label{eq5}
x(t) &=& \frac{x(0)}{2} \left[\cos\left\{\left(\sqrt{\omega^2 + \Omega^2/4} + \Omega/2\right)t\right\} + \cos\left\{\left(\sqrt{\omega^2 + \Omega^2/4} - \Omega/2\right)t\right\}\right]   \nonumber \\
&+& \frac{y(0)}{2} \left[\sin\left\{\left(\sqrt{\omega^2 + \Omega^2/4} + \Omega/2\right)t\right\} - \sin\left\{\left(\sqrt{\omega^2 + \Omega^2/4} - \Omega/2\right)t\right\}\right]   \;\;\;,
\end{eqnarray}

\noindent
and

\begin{eqnarray}\label{eq6}
y(t) &=& \frac{y(0)}{2} \left[\cos\left\{\left(\sqrt{\omega^2 + \Omega^2/4} + \Omega/2\right)t\right\} + \cos\left\{\left(\sqrt{\omega^2 + \Omega^2/4} - \Omega/2\right)t\right\}\right]   \nonumber \\
&-& \frac{x(0)}{2} \left[\sin\left\{\left(\sqrt{\omega^2 + \Omega^2/4} + \Omega/2\right)t\right\} - \sin\left\{\left(\sqrt{\omega^2 + \Omega^2/4} - \Omega/2\right)t\right\}\right]   \;\;\;.
\end{eqnarray}

\noindent
The above relations (\ref{eq5}-\ref{eq6}) satisfy the initial condition. For further check, one can show easily that Eq. (\ref{eq5}-\ref{eq6}) reduce to the expected results at the limit $\Omega = 0.0$. However, it is apparent in the above solutions that how the coupling of the two dimensional motion through the cross effect of the velocity dependent magnetic force  modifies the respective amplitudes.  Another important point is to be noted here that  $x(t)$ as well as $y(t)$ are superposition of periodic terms with periods

\begin{equation}\label{eq7}
T_1 = \frac{2 \pi}{\sqrt{\omega^2 + \Omega^2/4} + \Omega/2}   \;\;\;,
\end{equation}

\noindent
and

\begin{equation}\label{eq8}
T_2 = \frac{2 \pi}{\sqrt{\omega^2 + \Omega^2/4} - \Omega/2}   \;\;\;. 
\end{equation}

\noindent
Their ratio can be read as $\frac{T_2}{T_1} = \frac{\omega^2 + \Omega^2/2 + \Omega\sqrt{\omega^2 + \Omega^2/4}}{\omega^2}$. If the ratio is an integer  then motion would be a simple periodic one otherwise the dynamics may be quasi periodic in nature. One may obtain the condition for the integer ratio in the following way. Let $n \ge 1$ is a ratio between the two periods. Then we have

\begin{equation}\label{eq9}
\Omega^2 + 2 \Omega \sqrt{\omega^2 + \frac{\Omega^2}{4}} - 2 (n-1) \omega^2 = 0   \;\;\;.
\end{equation}

\noindent
The solution of the above equation can be read as

\begin{equation}\label{eq10}
\Omega = \frac{(n - 1) \omega}{\sqrt{n}}   \;\;\;.
\end{equation}

\begin{figure}[!htb]
\includegraphics[width=16cm,angle=0,clip]{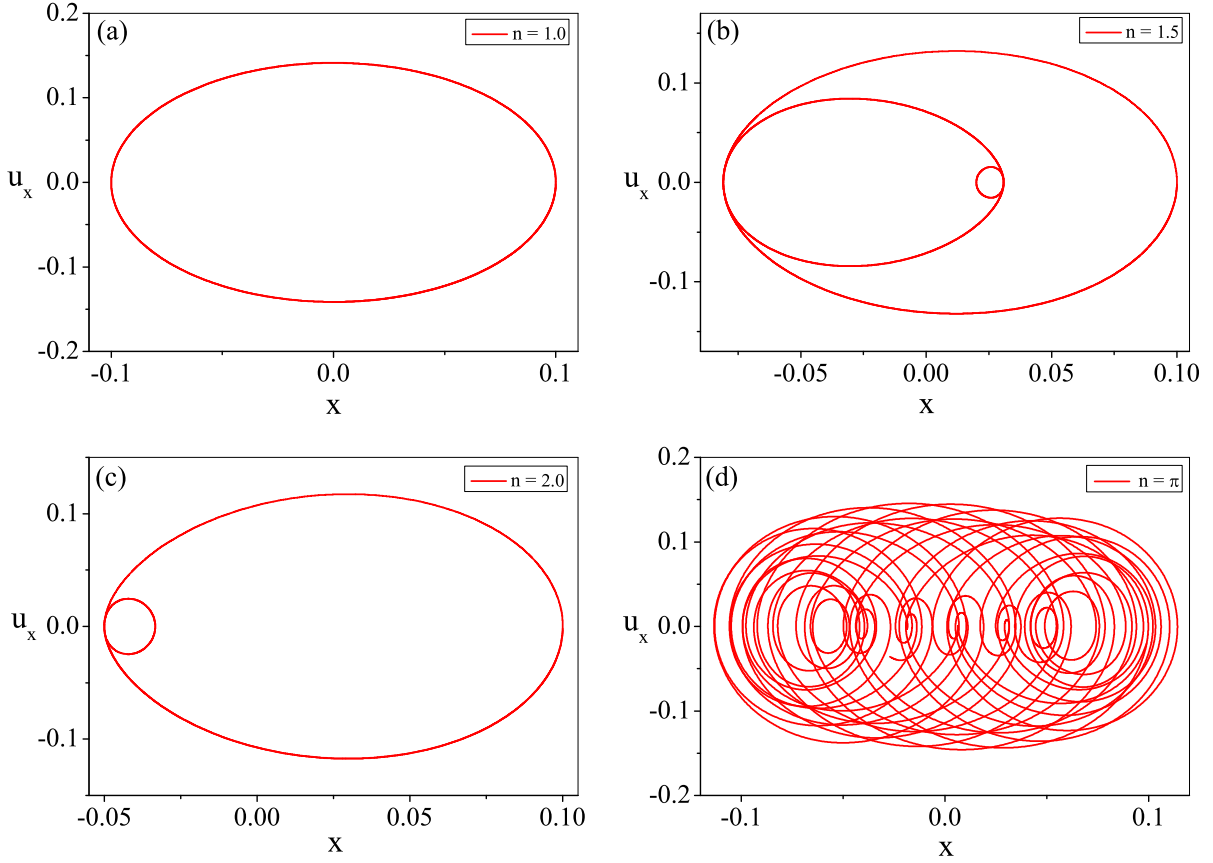}
\caption{{\footnotesize Plot of $u_x$ {\it vs.} $x$ for different values of n along with the relevant  parameter set,  $\omega^2 = 2.0, x(0) = 0.1, y(0) = 0.0, u_x (0) = 0.0$ and $u_y (0) = 0.0$ (Units are arbitrary)}}
\label{fig.1}
\end{figure}

\noindent
The above relation suggests that for $n = 1$, $\Omega = 0$. It is a check of the above calculation. For $\Omega = \omega$, $n$ is  $\frac{3 + \sqrt{5}}{2} \simeq 2.618033989....$. Thus the motion seems to be quasi periodic in nature for $\Omega = \omega$. In Fig.\ref{fig.1}, we have demonstrated the feature of the dynamics based on Eqs. (\ref{eq5}-\ref{eq6}) for different $\Omega$. It is fully consistent with Eq. (\ref{eq10}).

%\begin{figure}[!h]
%\centering
%\includegraphics[width=16 cm,height=11 cm]{figure_1.pdf}
%\caption{{\footnotesize Plot of $u_x$ vs. $x$ for different values of n along with the relevant  parameter set,  $\omega^2 = 2.0, x(0) = 0.1, y(0) = 0.0, u_x (0) = 0.0$ and $u_y (0) = 0.0$ (Units are arbitrary)}}
%\label{fig.1}
%\end{figure}

We now consider another condition, $\gamma = 0$ and $\omega = 0$. Then from Eq. (\ref{eq4}) we have

\begin{eqnarray}\label{eq11}
x(t) = \frac{x(0)}{2} \big[1+ \cos(\Omega t)\big] - \frac{y(0)}{2} \sin(\Omega t)   \;\;\;,
\end{eqnarray}

\noindent
and

\begin{eqnarray}\label{eq12}
y(t) = \frac{y(0)}{2} \big[1+ \cos(\Omega t)\big] - \frac{x(0)}{2} \sin(\Omega t)   \;\;\;.
\end{eqnarray}

\noindent
If $y(0) = 0$ then above equations (\ref{eq11}-\ref{eq12}) imply expected cyclotron motion with radius $= x(0)/2$, frequency $= \Omega$ and speed $= \{x(0) \Omega\}/2 = radius \times cyclotron frequency$. Similarly for $x(0) = 0$ Eqs.(\ref{eq11}-\ref{eq12}) also imply expected cyclotron motion. These are the checks of the present calculation. However, if both $x(0)$ and $y(0)$ are not zero then center of the cycle will oscillate in time along a line in the $x-y$ plane. Thus we may have a complicated cycloid type motion.

For further check, we consider the condition, $\Omega = 0$. Then  Eq. (\ref{eq4}) reduces to

\begin{eqnarray}\label{eq13}
x(t) = x(0) e^{- (\gamma t)/2} \cos\left(\sqrt{\omega^2 - \gamma^2/4} t\right)  \;\;\;,
\end{eqnarray}

\noindent
and

\begin{eqnarray}\label{eq14}
y(t) = y(0) e^{- (\gamma t)/2} \sin\left(\sqrt{\omega^2 - \gamma^2/4} t\right)  \;\;\;.
\end{eqnarray}

\noindent
These relations imply the expected damped oscillation at each direction. For $\gamma = 0$, the above relations reduce to the well known results. For direct check, one may arrive to these results putting $\Omega = 0$ and $\gamma = 0$ in Eq.(\ref{eq4}).

We now consider the approximate solutions at the limit, $\frac{B}{A} \rightarrow 0$, when both $\Omega$ and $\gamma$ may not be zero. Then from Eq. (\ref{eq4a}) we have

\begin{equation}\label{eq4b}
\theta \simeq \sqrt{A} \left(1 + i \frac{B}{2A}\right) t   \;\;\;.
\end{equation}

\noindent
Using this relation in Eq.(\ref{eq4}) one may have

\begin{eqnarray}\label{eq15}
x(t) &\simeq & \frac{x_0}{4} p_0 e^{- \frac{\gamma t}{2}} \left(e^{\frac{B t}{2 \sqrt{A}}} + e^{- \frac{B t}{2 \sqrt{A}}}\right) - \frac{x_0}{4} q_0 e^{- \frac{\gamma t}{2}} \left(e^{\frac{B t}{2 \sqrt{A}}} - e^{- \frac{B t}{2 \sqrt{A}}}\right)   \nonumber \\ \nonumber \\
&+& \frac{y_0}{4} s_0 e^{- \frac{\gamma t}{2}} \left(e^{\frac{B t}{2 \sqrt{A}}} - e^{- \frac{B t}{2 \sqrt{A}}}\right) - \frac{y_0}{4} r_0 e^{- \frac{\gamma t}{2}} \left(e^{\frac{B t}{2 \sqrt{A}}} - e^{- \frac{B t}{2 \sqrt{A}}}\right)    \;\;\;
\end{eqnarray}

\noindent
and

\begin{eqnarray}\label{eq16}
y(t) &\simeq & \frac{y_0}{2} p_0 e^{- \frac{\gamma t}{2}} \left(e^{\frac{B t}{2 \sqrt{A}}} + e^{- \frac{B t}{2 \sqrt{A}}}\right) - \frac{y_0}{4} q_0  e^{- \frac{\gamma t}{2}} \left(e^{\frac{B t}{2 \sqrt{A}}} - e^{- \frac{B t}{2 \sqrt{A}}}\right)   \nonumber \\ \nonumber \\
&+& \frac{x_0}{4} s_0  e^{- \frac{\gamma t}{2}} \left(e^{\frac{B t}{2 \sqrt{A}}} - e^{- \frac{B t}{2 \sqrt{A}}}\right) - \frac{x_0}{4} r_0 e^{- \frac{\gamma t}{2}} \left(e^{\frac{B t}{2 \sqrt{A}}} - e^{- \frac{B t}{2 \sqrt{A}}}\right)     \;\;\;
\end{eqnarray}

\noindent
where 
$p_0 = \cos \left\{\left(\sqrt{A} + \frac{\Omega}{2}\right)t\right\} + \cos \left\{\left(\sqrt{A} - \frac{\Omega}{2}\right)t\right\}$,
$q_0 = \cos \left\{\left(\sqrt{A} + \frac{\Omega}{2}\right)t\right\} - \cos \left\{\left(\sqrt{A} - \frac{\Omega}{2}\right)t\right\}$, $r_0 = \sin \left\{\left(\sqrt{A} + \frac{\Omega}{2}\right)t\right\} + \sin \left\{\left(\sqrt{A} - \frac{\Omega}{2}\right)t\right\}$ and $s_0 = \sin \left\{\left(\sqrt{A} + \frac{\Omega}{2}\right)t\right\} - \sin \left\{\left(\sqrt{A} - \frac{\Omega}{2}\right)t\right\}$.
The above equations imply the damped oscillation which is composed of two frequencies, $\sqrt{A} + \frac{\Omega}{2}$ and $\sqrt{A} - \frac{\Omega}{2}$, respectively. Now one can check easily that they reduce to all the exact results at the appropriate limits as mentioned above. Furthermore, from these approximate solutions one may infer the damped cyclotron motion for $A = \Omega^2/4 - \gamma^2/4$.

Before leaving this part we would mention that the argument in the sinusoidal function implies the condition at which the resonance may appear if the dynamical system (\ref{eq1}-\ref{eq2}) is driven periodically. In the next subsection we will check it based on the steady state dynamics.

\subsection{Steady state dynamics: Resonance condition and energetics}

We are now in a position to include an electric field (which is periodic in time with the angular frequency, $\omega_E$) in Eqs. (\ref{eq1}-\ref{eq2}) to explore the resonance condition and the related aspect. Let the electric field applied be as follows

\begin{equation}\label{elf}
{\bf E} = \hat{i} E_{0x} \cos(\omega_E t) + \hat{j} E_{0y} \cos(\omega_E t)   \;\;\;.
\end{equation}

\noindent
Here $E_x(t) = E_{0x} \cos (\omega_E t)$ and $E_y(t) = E_{0y} \cos (\omega_E t)$ are the components of the applied electric field. Then the Eqs. (\ref{eq1}-\ref{eq2}) of motion become

\begin{equation}\label{eqr1}
m \ddot{x} = - m \omega^2 x - m \gamma \dot{x} + m \Omega \dot{y} + q E_{0x} \cos(\omega_E t)   \;\;\;,
\end{equation}

\noindent
and

\begin{equation}\label{eqr2}
m \ddot{y} = - m \omega^2 x - m \gamma \dot{y} - m \Omega \dot{x} + q E_{0y} \cos(\omega_E t)   \;\;\;.
\end{equation}

\noindent
Thus the components of driving force are $F_x = q E_{0x} \cos(\omega_E t)$ and $F_y = q E_{0y} \cos(\omega_E t)$, respectively. One may now choose the following particular solutions for the above equations

\begin{equation}\label{eqr3}
x(t) = a \cos(\omega_E t - \phi_1)   \;\;\;,
\end{equation}

\noindent
and

\begin{equation}\label{eqr4}
y(t) = b \cos(\omega_E t - \phi_2)   \;\;\;. 
\end{equation}

\noindent
Here $\phi_1$ and $\phi_2$ are two relevant phase constants. Using relations (\ref{eqr3}-\ref{eqr4}) into Eqs. (\ref{eqr1}-\ref{eqr2}) we have

\begin{eqnarray}\label{eqr5}
&& \left\{a \left(\omega^2 - \omega_E^2\right) \cos \phi_1\right\} \cos(\omega_E t) + \left\{a \left(\omega^2 - \omega_E^2\right) \sin \phi_1\right\} \sin(\omega_E t)   \nonumber  \\
&=& (a \gamma \omega_E \cos \phi_1) \sin(\omega_E t) - (a \gamma \omega_E \sin \phi_1) \cos(\omega_E t) - (b \Omega \omega_E \cos \phi_2) \sin(\omega_E t)   \nonumber  \\
&+& (b \Omega \omega_E \sin \phi_2) \cos(\omega_E t) + \frac{q}{m} E_{0x} \cos(\omega_E t)   \;\;\;.
\end{eqnarray}

\noindent
and

\begin{eqnarray}\label{eqr6}
&& \left\{b \left(\omega^2 - \omega_E^2\right) \cos \phi_2\right\} \cos(\omega_E t) + \left\{b \left(\omega^2 - \omega_E^2\right) \sin \phi_2\right\} \sin(\omega_E t)   \nonumber  \\
&=& (b \gamma \omega_E \cos \phi_2) \sin(\omega_E t) - (b \gamma \omega_E \sin \phi_2) \cos(\omega_E t) + (a \Omega \omega_E \cos \phi_1) \sin(\omega_E t)   \nonumber  \\
&-& (a \Omega \omega_E \sin \phi_1) \cos(\omega_E t) + \frac{q}{m} E_{0y} \cos(\omega_E t)   \;\;\;.
\end{eqnarray}

\noindent
Comparing the coefficients of $\cos(\omega_E t)$ and $\sin(\omega_E t)$ in the both sides of Eq. (\ref{eqr5}) we get the following relations

\begin{equation}\label{eqr7}
a \left(\omega^2 - \omega_E^2\right) \cos \phi_1 = \frac{q}{m} E_{0x} - a \gamma \omega_E \sin \phi_1 + b \Omega \omega_E \sin \phi_2   \;\;\;,
\end{equation}

\noindent
and

\begin{equation}\label{eqr8}
a \left(\omega^2 - \omega_E^2\right) \sin \phi_1 = a \gamma \omega_E \cos \phi_1 - b \Omega \omega_E \cos \phi_2   \;\;\;.
\end{equation}

\noindent
Similarly from Eq. (\ref{eqr6}) we get the following relations

\begin{equation}\label{eqr9}
b \left(\omega^2 - \omega_E^2\right) \cos \phi_2 = \frac{q}{m} E_{0y} - b \gamma \omega_E \sin \phi_2 - a \Omega \omega_E \sin \phi_1   \;\;\;,
\end{equation}

\noindent
and

\begin{equation}\label{eqr10}
b \left(\omega^2 - \omega_E^2\right) \sin \phi_2 = b \gamma \omega_E \cos \phi_2 + a \Omega \omega_E \cos \phi_1   \;\;\;.
\end{equation}

\noindent
Now from coupled Eqs. (\ref{eqr7}-\ref{eqr10}) we have

\begin{equation}\label{eqr11}
a = \frac{\sqrt{H_1^2 + H_2^2}}{H_0}   \;\;\;,
\end{equation}

\begin{equation}\label{eqr12}
b = \frac{\sqrt{H_3^2 + H_4^2}}{H_0}   \;\;\;,
\end{equation}

\begin{equation}\label{eqr13}
\tan \phi_1 = \frac{H_2}{H_1}   \;\;\;.
\end{equation}

\noindent
and

\begin{equation}\label{eqr14}
\tan \phi_2 = \frac{H_4}{H_3}   \;\;\;.
\end{equation}

\noindent
Here we have used

\begin{equation}\label{eqr15}
H_0 = \left\{\left(\omega^2 - \omega_E^2\right)^2 - \left(\Omega^2 - \gamma^2\right) \omega_E^2\right\}^2 + 4 \gamma^2 \Omega^2 \omega_E^4   \;\;\;,
\end{equation}

\begin{equation}\label{eqr16}
H_1 = \frac{q}{m} \left(\omega^2 - \omega_E^2\right) \left[\left\{\left(\omega^2 - \omega_E^2\right)^2 - \left(\Omega^2 - \gamma^2\right) \omega_E^2\right\} E_{0x} + 2 \gamma \Omega \omega_E^2 E_{0y} \right]   \;\;\;,
\end{equation}

\begin{equation}\label{eqr17}
H_2 = \frac{q}{m} \omega_E \left[\left\{\left(\omega^2 - \omega_E^2\right)^2 - \left(\Omega^2 - \gamma^2\right) \omega_E^2\right\} \left(\gamma E_{0x} - \Omega E_{0y}\right) + 2 \gamma \Omega \omega_E^2 \left(\gamma E_{0y} + \Omega E_{0x}\right)\right]   \;\;\;,
\end{equation}

\begin{equation}\label{eqr18}
H_3 = \frac{q}{m} \left(\omega^2 - \omega_E^2\right) \left[\left\{\left(\omega^2 - \omega_E^2\right)^2 - \left(\Omega^2 - \gamma^2\right) \omega_E^2\right\} E_{0y} - 2 \gamma \Omega \omega_E^2 E_{0x} \right]   \;\;\;,
\end{equation}

\noindent
and

\begin{equation}\label{eqr19}
H_4 = \frac{q}{m} \omega_E \left[\left\{\left(\omega^2 - \omega_E^2\right)^2 - \left(\Omega^2 - \gamma^2\right) \omega_E^2\right\} \left(\gamma E_{0y} + \Omega E_{0x}\right) - 2 \gamma \Omega \omega_E^2 \left(\gamma E_{0x} - \Omega E_{0y}\right)\right]   \;\;\;.
\end{equation}

\noindent
Thus the amplitudes of the particular solutions carry the interesting signature of the magnetic field induced coupling of the two dimensional motion through the similarity in their structures in terms of the relevant parameters, frequencies of the harmonic oscillator and the driving field, amplitudes of the driving fields and strength of the magnetic field, respectively. $a$ becomes $b$ on replacement of $\Omega$ by $- \Omega$ and interchange of position between $E_{0x}$ and $E_{0y}$. Thus even for $E_{0x} = E_{0y}$, $a \neq b$. It is a signature of the magnetic force induced breakdown of the equivalence between the motions along the two directions. Eqs. (\ref{eqr11}-\ref{eqr14}) imply the role of another velocity dependent dissipative force in this context. Similarly one may expect the signature of the breakdown of the equivalence in terms of the phase shift between input and output signals. Shortly we will discuss this in detail.

We are now in a position to dertermine the resonance conditions. It seems to be difficult to find the resonance condition since the derivative of $a$ or $b$ with respect to the driving frequency may correspond to an algebraic equation which is not solvable  analytically. Before going to predict the approximate resonance condition, we show that Eqs. (\ref{eqr11}-\ref{eqr14}) reduce to the well known results for $\Omega = 0$. At this limit Eqs. (\ref{eqr15}-\ref{eqr19}) become

\begin{equation}\label{eqr20}
H_0 = \left\{\left(\omega^2 - \omega_E^2\right)^2 + \gamma^2 \omega_E^2\right\}^2   \;\;\;,
\end{equation}

\begin{equation}\label{eqr21}
H_1 = \frac{q}{m} E_{0x} \left(\omega^2 - \omega_E^2\right) \left\{\left(\omega^2 - \omega_E^2\right)^2 + \gamma^2 \omega_E^2\right\}   \;\;\;,
\end{equation}

\begin{equation}\label{eqr22}
H_2 = \frac{q}{m} E_{0x} \gamma \omega_E \left\{\left(\omega^2 - \omega_E^2\right)^2 + \gamma^2 \omega_E^2\right\}   \;\;\;,
\end{equation}

\begin{equation}\label{eq23}
H_3 = \frac{q}{m} E_{0y} \left(\omega^2 - \omega_E^2\right) \left\{\left(\omega^2 - \omega_E^2\right)^2 + \gamma^2 \omega_E^2\right\}   \;\;\;,
\end{equation}

\noindent
and

\begin{equation}\label{eqr24}
H_4 = \frac{q}{m} E_{0y} \gamma \omega_E \left\{\left(\omega^2 - \omega_E^2\right)^2 + \gamma^2 \omega_E^2\right\}   \;\;\;.
\end{equation}

Then Eqs.(\ref{eqr11}-\ref{eqr14}) become

\begin{equation}\label{eqr25}
a =  \frac{q E_{0x}}{m \left[\left(\omega^2 - \omega_E^2\right)^2 + \gamma^2 \omega_E^2\right]^{1/2}}   \;\;\;,
\end{equation}

\begin{equation}\label{eqr26}
b =  \frac{q E_{0y}}{m \left[\left(\omega^2 - \omega_E^2\right)^2 + \gamma^2 \omega_E^2\right]^{1/2}}   \;\;\;,
\end{equation}
\begin{equation}\label{eqr27}
\tan \phi_1 = \tan \phi_2 = \frac{\gamma \omega_E}{\omega^2 - \omega_E^2}   \;\;\;.
\end{equation}

\noindent
Eqs. (\ref{eqr25}-\ref{eqr27}) imply a very good check of the present calculation. For further check, one may determine easily the following resonance condition from Eqs. (\ref{eqr25}-\ref{eqr26}),

\begin{equation}\label{eqr29}
\omega_E = \omega \left(1 - \frac{\gamma^2}{2 \omega^2}\right)^{1/2}   \;\;\;.
\end{equation}

\noindent
This condition is implied by the transient motion (\ref{eq13}-\ref{eq14}) in the presence of the driving force.  It reduces to another well known result, $\omega_E = \omega$ for $\gamma = 0$. Then $a = \frac{q E_{0x}}{m \left(\omega^2 - \omega_E^2\right)}$, $b = \frac{q E_{0y}}{m \left(\omega^2 - \omega_E^2\right)}$ and $\phi_1 = \phi_2 = 0$. For the direct check, one may arrive to these results using $\gamma = \Omega = 0$ in Eqs. (\ref{eqr11}-\ref{eqr14}). However, comparing Eqs. (\ref{eqr26}-\ref{eqr27}) we find a contrast result in the presence of a magnetic field. The effect of dissipative force may  also appear in the numerator of the amplitude function as a signature of the modulation of the frequency of the dynamics by the field. In other words, the interference between the two driving components through velocity dependent coupling results to appear $\gamma$, $\Omega$, $E_{0x}$ and $E_{0y}$ in the numerator of the amplitude functions, $a$ and $b$, respectively. 

Now we have to determine the resonance condition in the presence of a magnetic field. One may determine it applying a trick by inspection of both numerator and denominator in Eqs. (\ref{eqr11}-\ref{eqr12}). At the weak damping limit when the resonance phenomenon may  appear then $H_0$ may be minimum around the following condition,

\begin{equation}\label{eqr30}
\left(\omega^2 - \omega_E^2\right)^2 - \Omega^2 \omega_E^2 = 0   \;\;\;,
\end{equation}

\noindent
Thus the approximate resonance conditions may be read as

\begin{equation}\label{eqr31}
\omega_L \simeq \sqrt{\omega^2 + \frac{\Omega^2}{4}} - \frac{\Omega}{2}   \;\;\;,
\end{equation}

\noindent
and

\begin{equation}\label{eqr32}
\omega_R \simeq \sqrt{\omega^2 + \frac{\Omega^2}{4}} + \frac{\Omega}{2}   \;\;\;.
\end{equation}

\begin{figure}[!htb]
\includegraphics[width=16cm,angle=0,clip]{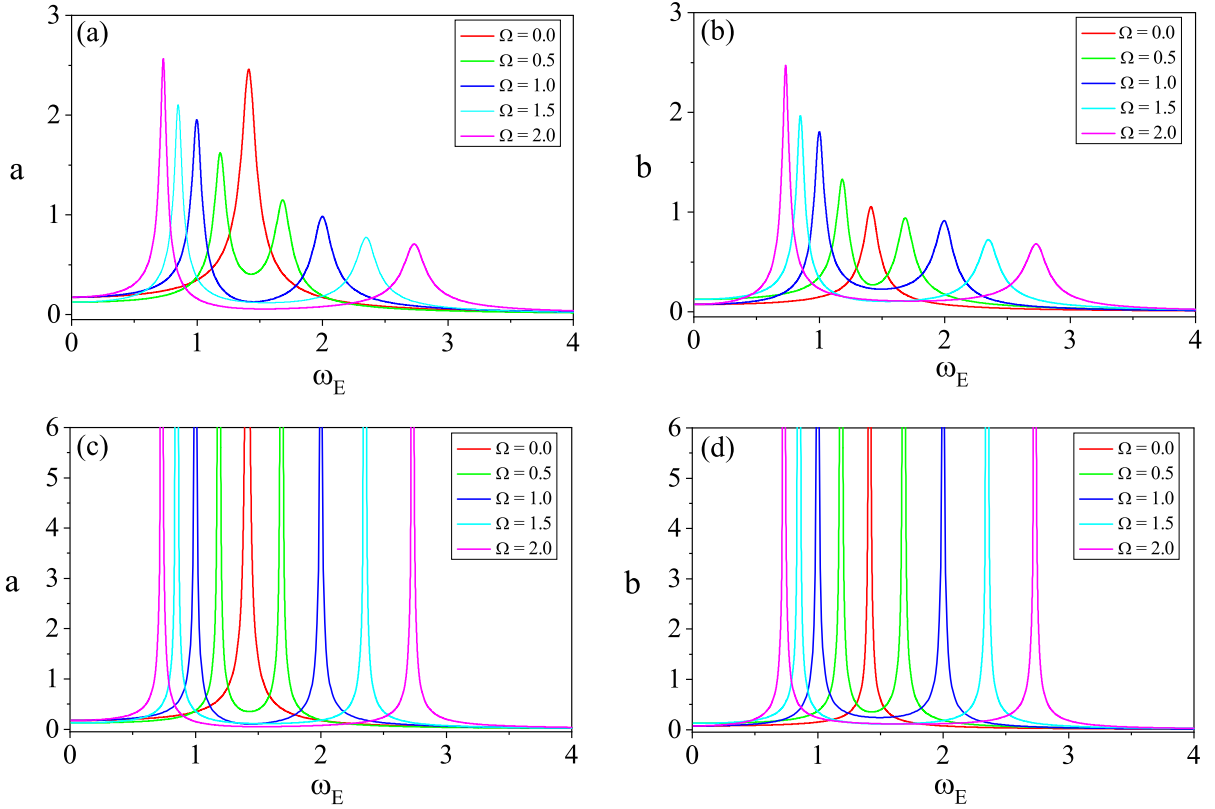}
\caption{{\footnotesize Plot of $a$ and $b$ vs. $\omega_E$ for different values of $\Omega$. (a) $\omega^2 = 2.0, \gamma = 0.1$ and $ E_{0x} = E_{0y} = 0.25$ (b) $\omega^2 = 2.0, \gamma = 0.1$ and $ E_{0x} = E_{0y} = 0.25$ (c) $\omega^2 = 2.0, \gamma = 0.0$ and $E_{0x} = E_{0y} = 0.25$ (d) $\omega^2 = 2.0, \gamma = 0.0$ and $ E_{0x} = E_{0y} = 0.25$ (Units are arbitrary)}}
\label{fig.3}
\end{figure}

\noindent
It is to be noted here that the above conditions are very closed to those which are implied in Eqs. (\ref{eq15}-\ref{eq16}). To check the validity of our calculation we have demonstrated the exact results (\ref{eqr11}-\ref{eqr12}) in Fig.\ref{fig.3}.

\begin{table}[ht]
\caption{Comparison between theoretically calculated resonating frequencies and the exact result for the driven damped two dimensional harmonic oscillator}
\begin{center}
\begin{tabular}{|c|c|c|c|c|}
\hline
Value of &
\multicolumn{2}{|c|}{Resonance at $\omega_L$} &
\multicolumn{2}{|c|}{Resonance at $\omega_R$} \\
\cline{2-5}
$\Omega$ & Theoretical & Exact & Theoretical & Exact \\ 
\hline
0.5 & 1.190 & 1.189 & 1.680 & 1.679 \\
\hline
1.0 & 1.002 & 1.000 & 1.997 & 1.996 \\
\hline
1.5 & 0.852 & 0.849 & 2.348 & 2.347 \\
\hline
2.0 & 0.732 & 0.729 & 2.730 & 2.730 \\
\hline
\end{tabular}
\end{center}
\label{tab.1}
\end{table}

\noindent
The resonance conditions according to this figure are compared with the analytically calculated results in Table \ref{tab.1}. It shows that  there is a very good agreement between the theoretical and the exact results. Another important point is to be noted here that the Fig.\ref{fig.3} exhibits an asymmetric splitting of the spectrum with the output signal. The splitting is implied by the solutions (\ref{eq15}-\ref{eq16}) which are composed of two frequencies. Following the above approximation, the amplitudes at the resonance condition may be read as

\begin{equation}\label{eqr33}
a \simeq \frac{q}{m}\frac{\sqrt{\left(\omega^2 - \omega_E^2\right)^2 \left(\gamma E_{0x} + 2 \Omega E_{0y} \right)^2+\omega_E^2 \left[\gamma\left(\gamma E_{0x} - \Omega E_{0y}\right) + 2 \Omega \left(\gamma E_{0y} + \Omega E_{0x}\right)\right]^2}}{\gamma^3 \omega_E^2+ 4\gamma \Omega^2 \omega_E^2}   \;\;\;,
\end{equation}

\noindent
and

\begin{equation}\label{eqr34}
b \simeq \frac{q}{m}\frac{\sqrt{\left(\omega^2 - \omega_E^2\right)^2 \left(\gamma E_{0y} - 2 \gamma \Omega E_{0x} \right)^2+\omega_E^2 \left[\gamma \left(\gamma E_{0y} + \Omega E_{0x}\right) - 2 \gamma \Omega \left(\gamma E_{0x} - \Omega E_{0y}\right)\right]^2}}{\gamma^3 \omega_E^2 + 4\gamma \Omega^2 \omega_E^2}   \;\;\;,
\end{equation}

\noindent
The above equations imply that at the resonance condition, the amplitude may decreases with increase in the resonating frequency. In panels (a) and (b) of Fig.\ref{fig.3} we have demonstrated nature of the asymmetric splitting for different strength of the magnetic field. 

We now check the fate of the asymmetric splitting in the absence of damping. In the absence of dissipative force, Eqs. (\ref{eqr11}-\ref{eqr14}) become

\begin{equation}\label{eqr35}
a = \frac{q \sqrt{E_{0x}^2 \left(\omega^2 - \omega_E^2\right)^2 + E_{0y}^2 \Omega^2 \omega_E^2}}{m \left\{\left(\omega^2 - \omega_E^2\right)^2 - \Omega^2 \omega_E^2\right\}}   \;\;\;,
\end{equation}

\begin{equation}\label{eqr36}
b = \frac{q \sqrt{E_{0y}^2 \left(\omega^2 - \omega_E^2\right)^2 + E_{0x}^2 \Omega^2 \omega_E^2}}{m \left\{\left(\omega^2 - \omega_E^2\right)^2 - \Omega^2 \omega_E^2\right\}}    \;\;\;,
\end{equation}

\begin{equation}\label{eqr37}
\tan\phi_1 = - \frac{E_{0y} \Omega \omega_E}{E_{0x} (\omega^2 - \omega_E^2)}  
\end{equation}

\noindent
and

\begin{equation}\label{eqr38}
\tan\phi_2 = \frac{E_{0x} \Omega \omega_E}{E_{0y} (\omega^2 - \omega_E^2)}   \;\;\;.
\end{equation}

\noindent
The denominator of the amplitude functions imply that the resonating frequencies may be equal to those which are given by Eqs. (\ref{eqr31}-\ref{eqr32}). These are consistent with Eqs. (\ref{eq5}-\ref{eq6}). Then amplitude becomes infinity as suggested by Eqs. (\ref{eqr33}-\ref{eqr34}). To check this we have demonstrated Eqs. (\ref{eqr11}-\ref{eqr12}) in panels (c) and (d) of Fig.\ref{fig.3}. It exactly corresponds to Eqs. (\ref{eqr33}-\ref{eqr34}). Thus in the absence of damping, the asymmetric nature of the splitting is not clear as like as the previous case.
 
 \begin{figure}[!htb]
\includegraphics[width=16cm,angle=0,clip]{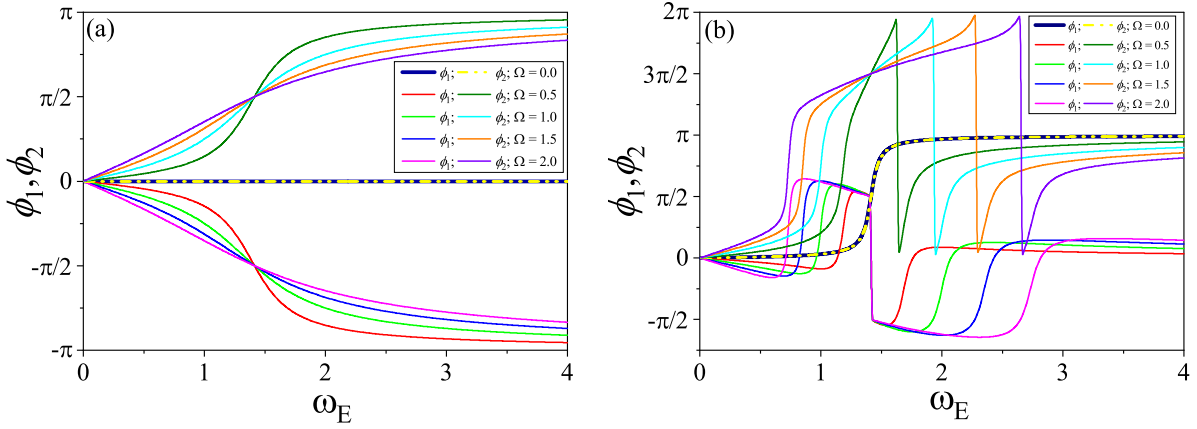}
\caption{{\footnotesize Plot of $\phi_1$ and $\phi_2$ vs. $\omega_E$ for different values of $\Omega$ along with the relevant  parameter set, $\omega^2 = 2.0$ and $E_{0x} = E_{0y} = 0.25$. (a) $\gamma = 0.0$ (b) $\gamma = 0.1$ (Units are arbitrary).}}
\label{fig.2}
\end{figure}

We now consider the phase shift between the input and output signals.
Using Eqs. (\ref{eqr13}-\ref{eqr14}), the signature of the breakdown of the equivalence in terms of the phase shift between the input and the output signal for the respective directions has been demonstrated in Fig.\ref{fig.2} both for dissipative and non-dissipative systems. Panel (a) of this figure shows that the magnetic field induces the phase shift even in the absence of damping. Here the magnetic force takes the similar role as like as the dissipative one \cite{feyn, symon}. This is explicit in Eqs.(\ref{eqr37}-\ref{eqr38}). Thus the phase shift at $\omega=\omega_E$, may not depend on the strength of the applied magnetic field as a signature of similarity between damping strength and MF. Now one may address the difference between these as we expect from the equations of motion.
The relevant points are to be noted here. First, the phase shift may depend on the amplitude of the driving force in the presence of the magnetic field. It is a sharp contrast behavior compared to the case where the damping induces the phase shift which is independent on the amplitude of the driving force (\ref{eqr27}). Eq. (\ref{eqr37}) implies that the phase shift for the $x$-component motion is enhanced by the motion of the other direction in the presence of opposition from the driving force along the $x$-direction. Similarly, one may interpret Eq. (\ref{eqr38}). Thus here the origin of the phase shift is the effect of the cross coupling which may mimic the role of dissipative action in the same context. Then it is expected that in the presence of damping force, the phase shift may depend on the amplitude of the periodic electric field in a more complicated way  as implied in Eqs. (\ref{eqr13}-\ref{eqr14}) as well as panel (b) of Fig.\ref{fig.2}. One may notice here the nature of change of the phase shits around the resonance conditions. It may be useful to corroborate the appearance of the  magnetic field induced anti resonance phenomenon. Shortly we will consider this issue.
Second, although there is a phase difference (at the resonance condition in the absence of damping force) between the input and outputs as like as the driven damped harmonic oscillator but the amplitudes of the output signal is still infinite as shown in panels (c) and (d) of Fig.\ref{fig.3}. The nature of divergence is quite similar to the driven harmonic oscillator in the absence of damping when the phase shift is zero. 
Thus the finite amplitude at the resonance condition in the presence of damping for the driven harmonic oscillator may be due to the dissipation of energy and here the phase shift may not have any role. It is an indirect conclusion with the help of driven harmonic oscillator in the presence of a magnetic field. Here the non zero phase shift gives  the indication that the resonating frequency may be different from the periodically driven harmonic oscillator. However,
as the amplitudes becomes infinite at the resonance condition for this case, the asymmetric nature (as expected from the numerator of the amplitude of the output signal) of the splitting is not prominent in the panels (c) and (d) as like as the other panels in the same figure. Thus the terms, $\gamma^2 \omega_E^2$ and $4 \gamma^2 \Omega^2 \omega_E^4$ (which appears in the denominator of the amplitude functions (\ref{eqr11}-\ref{eqr12}) due to dissipative action) are the leading quantities to modulated the nature of the asymmetric splitting of the spectrum of the output signal. In other words, one may observe a single peak instead of two at relatively high strength of the applied magnetic field as implied in panel (a) of  Fig.\ref{fig.3}.

\begin{figure}[!htb]
\includegraphics[width=16cm,angle=0,clip]{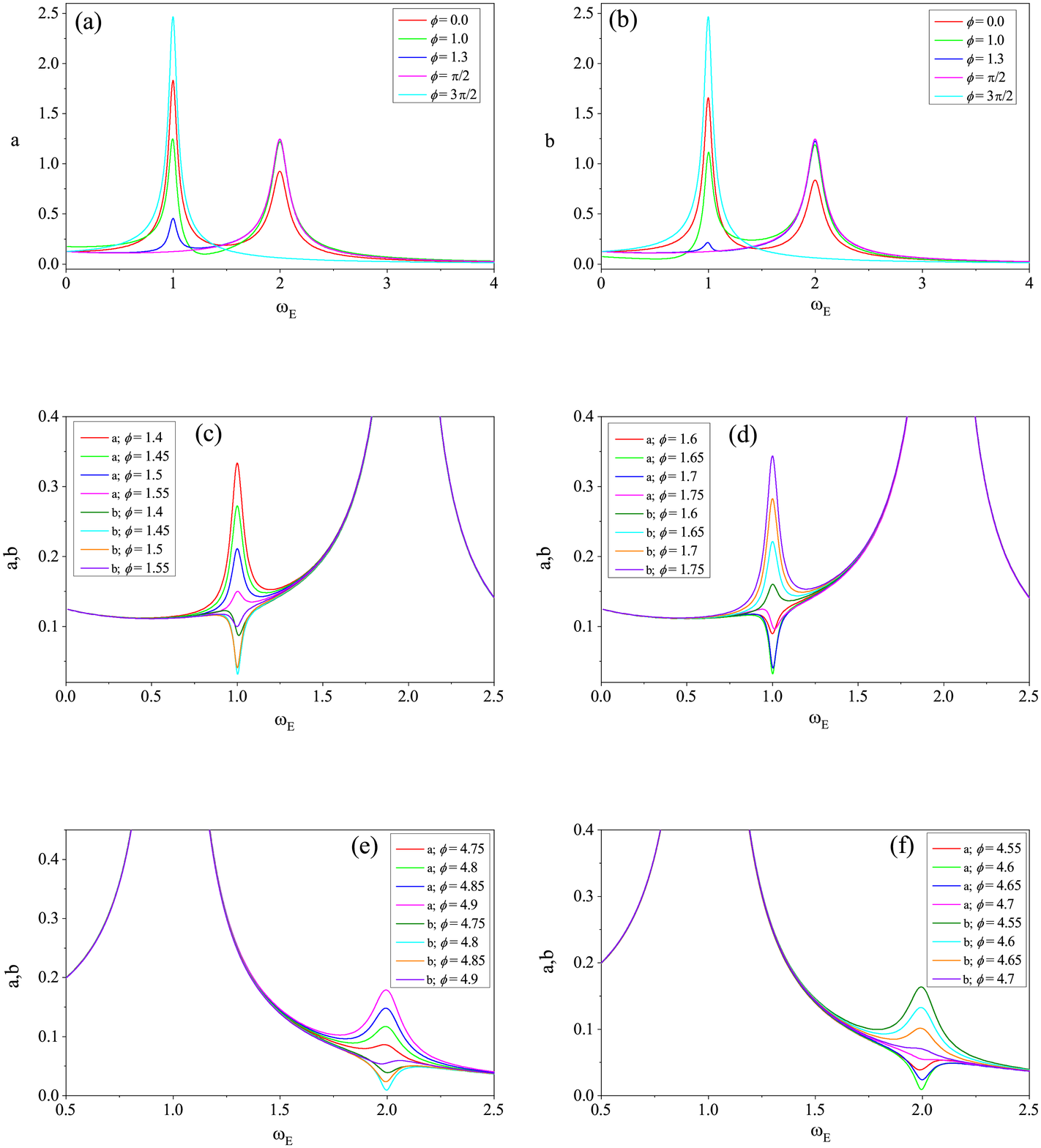}
\caption{{\footnotesize Plot of $a$ and $b$ vs. $\omega_E$ for different values of $\phi$ along with the parameter set,  $\omega^2 = 2.0, \gamma = 0.1$ and $E_{0x} = E_{0y} = 0.25$ (Units are arbitrary).}}
\label{fig.4}
\end{figure}

\subsection{Phase difference between the input signals: Anti resonance}

In the earlier discussion we have noted that the interference between the two driving components through the velocity dependent coupling may have important consequences. Then we consider the phase difference between the two input signals with the following driving electric field,

\begin{equation}\label{ph1}
{\bf E} = \hat{i} E_{0x} \cos(\omega_E t - \phi) + \hat{j} E_{0y} \cos(\omega_E t)   \;\;\;.
\end{equation}

\noindent
Then equations (\ref{eqr1}-\ref{eqr2}) of motion become

\begin{equation}\label{ph2}
m \ddot{x} = - m \omega^2 x - m \gamma \dot{x} + m \Omega \dot{y} + q E_{0x} \cos(\omega_E t - \phi)   \;\;\;,
\end{equation}

\noindent
and

\begin{equation}\label{ph3}
m \ddot{y} = - m \omega^2 x - m \gamma \dot{y} - m \Omega \dot{x} + q E_{0y} \cos(\omega_E t)   \;\;\;.
\end{equation}

\noindent
Choosing the particular solutions as like as given by Eqs. (\ref{eqr3}-\ref{eqr4}) and  following the above procedure we have

\begin{equation}\label{ph4}
a = \frac{\sqrt{H_1^2 + H_2^2}}{H_0}   \;\;\;,
\end{equation}

\begin{equation}\label{ph5}
b = \frac{\sqrt{H_3^2 + H_4^2}}{H_0}   \;\;\;,
\end{equation}

\begin{equation}\label{ph6}
\tan \phi_1 = \frac{H_2}{H_1} 
\end{equation}

and

\begin{equation}\label{ph7}
\tan \phi_2 = \frac{H_4}{H_3}   \;\;\;.
\end{equation}

\noindent
It is to be noted here that $H_0$ remains same as given by Eq. (\ref{eqr15}) but Eqs. (\ref{eqr16}-\ref{eqr19}) are modified as

\begin{eqnarray}\label{ph8}
H_1 = \frac{q}{m} E_{0x} \Big[\left(\omega^2 - \omega_E^2\right)^2 &-& \left(\Omega^2 - \gamma^2\right) \omega_E^2\Big] \left\{\left(\omega^2 - \omega_E^2\right) \cos \phi - \gamma \omega_E \sin \phi\right\}   \nonumber\\
&+& 2 \frac{q}{m} \gamma \Omega \omega_E^2 \left\{\left(\omega^2 - \omega_E^2\right) E_{0y} - E_{0x} \Omega \omega_E \sin \phi\right\}   \;\;\;,
\end{eqnarray}

\begin{equation}\label{ph9}
H_2 = \frac{1}{\left(\omega^2 - \omega_E^2\right)} \left(\frac{q}{m} E_{0x} H_0 \sin \phi + \gamma \omega_E H_1 - \Omega \omega_E H_3\right)   \;\;\;,
\end{equation}

\begin{eqnarray}\label{ph10}
H_3 = \frac{q}{m} \Big[\left(\omega^2 - \omega_E^2\right)^2 &-& \left(\Omega^2 - \gamma^2\right) \omega_E^2\Big] \left\{\left(\omega^2 - \omega_E^2\right) E_{0y} - E_{0x} \Omega \omega_E \sin \phi\right\}   \nonumber\\
&-& 2 \frac{q}{m} E_{0x} \gamma \Omega \omega_E^2 \left\{\left(\omega^2 - \omega_E^2\right) \cos \phi - \gamma \omega_E \sin \phi\right\}   \;\;\;
\end{eqnarray}

\noindent
and

\begin{equation}\label{ph11}
H_4 = \frac{\omega_E}{\left(\omega^2 - \omega_E^2\right)} \left(\gamma H_3 + \Omega H_1\right)   \;\;\;.
\end{equation}

\noindent
Putting $\phi = 0$ in Eqs. (\ref{ph8})-(\ref{ph11}) one  can show easily that Eqs. (\ref{ph4}-\ref{ph7}) reduce to  Eqs. (\ref{eqr11}-\ref{eqr14}). It constitutes an important check of the present calculation.

\begin{figure}[!htb]
\includegraphics[width=16cm,angle=0,clip]{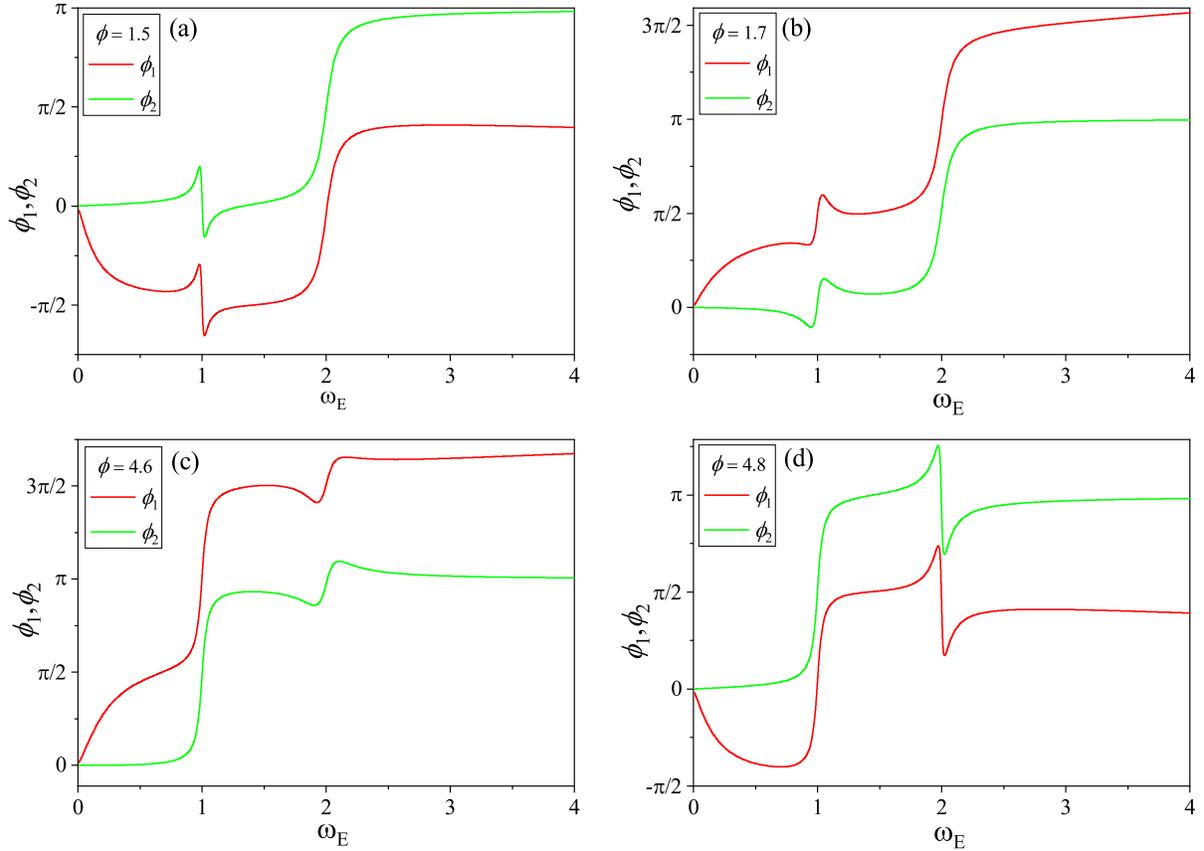}
\caption{{\footnotesize Plot of $\phi_1$ and $\phi_2$ vs. $\omega_E$ for different values of $\phi$ along with the relevant  parameter set, $\omega^2 = 2.0$, $\gamma = 0.1$, $\Omega = 1.0$ and $E_{0x} = E_{0y} = 0.25$. (Units are arbitrary).}}
\label{fig.4a}
\end{figure}

\noindent
It is really difficult to find out the resonant condition from Eqs. (\ref{ph4}-\ref{ph5}). But one may anticipate it by inspection of both numerator and denominator in Eqs. (\ref{ph4}-\ref{ph5}) as before. The condition would be remain same as that of $\phi = 0$ case since transients are same for both the situations. As a matter of fact same denominator ($H_0$) appears in Eqs. (\ref{eqr11}-\ref{eqr12}) and Eqs. (\ref{ph4}-\ref{ph5}), respectively. Thus the phase difference between the input signals may modulate amplitude of the output signals (\ref{ph4}-\ref{ph5}) without affecting the resonance condition as a signature of their interference through the velocity dependent coupling. This has been demonstrated in panels (a) and (b) of Fig.\ref{fig.4}. Another point is to be noted from this figure.
Comparing it with Fig.\ref{fig.3} we find that in the asymmetric splitting process, the peak height at higher resonating frequency may be greater than that of the other. An extreme case such as one of the peaks may disappear as shown in panels (a) and (b). Finally, the remaining panels (c) to (f) exhibits that the magnetic field may induce anti resonance phenomenon. Thus for a certain phase difference between the input signals the reduced amplitude at anti resonance can be regarded as due to destructive interference or cancellation of forces acting on the oscillator. In this context we have demonstrated variation of the relevant phase constants $\phi_1$ and $\phi_2$ with the driving frequency in Fig.\ref{fig.4a}. Comparing it with Fig.\ref{fig.2} we find that the nature of change of phase constants around the resonating driving frequency corroborates the appearance of the anti resonance. 
It is to be noted here that both the resonance and the anti resonance phenomena may appear at the same driving frequency depending on the phase difference between the input signals. This is consistent with the transient motion as well as Fig.\ref{fig.3}. As the anti resonance phenomenon is driven by the phase difference between the input signals then one may expect to appear it at the relevant frequency which may correspond to a normal mode. This is sharp contrast to the case of coupled oscillators having coupling through the conservative force field\cite{antir}. As a signature of this kind of coupling the system may constitute one or more frequencies which are different from the relevant normal mode(s) and anti resonance may appear at these frequencies. For this case resonance and anti resonance may appear alternatively in the output signal. Thus the magnetic field induced anti resonance is a distinct one. It is to be noted here that
there is a similarity between the two case. In the case of velocity dependent coupling, if a resonating frequency become anti resonating one then another frequency behaves as a resonating one.

\subsubsection{The energetics}

Although the magnetic force does not work but it may modulate the dynamics of a dynamical system through the modification of the characteristic frequency of the system as implied in the transient dynamics. Then the work done on the system by the driving force may depend on it. An indication regarding this already we have shown through the investigation on amplitude of the oscillation at the steady state. Thus we are in a position to calculate the power ($P$) due to the work done by the driving force \cite{feyn, symon}. It is the product between the force and the velocity. For the two dimensional motion the power is given by

\begin{equation}\label{en1}
P = F_x \frac{dx}{dt} + F_y \frac{dy}{dt}   \;\;\;.
\end{equation}

\noindent
It is an oscillating quantity. Then one may be interested to know the average power, ($\langle P \rangle$). Using Eqs. (\ref{eqr3}-\ref{eqr4}) into the above equation we obtain

\begin{equation}\label{en2}
\langle P \rangle = \frac{1}{2} m \gamma \omega_E^2 \left(a^2 + b^2\right)   \;\;\;.
\end{equation}

\noindent
Similarly one can determine the average energy as

\begin{equation}\label{en3}
\langle U \rangle = \frac{1}{2} m \omega_E^2 \frac{1}{2} \left(a^2 + b^2\right) + \frac{1}{2} m \omega^2 \frac{1}{2} \left(a^2 + b^2\right)      \;\;\;.
\end{equation}

\noindent
Eq. (\ref{eqr3}) implies that the first term corresponds to the stored mean potential energy and the other one is due to the average kinetic energy.

\begin{figure}[!htb]
\includegraphics[width=16cm,angle=0,clip]{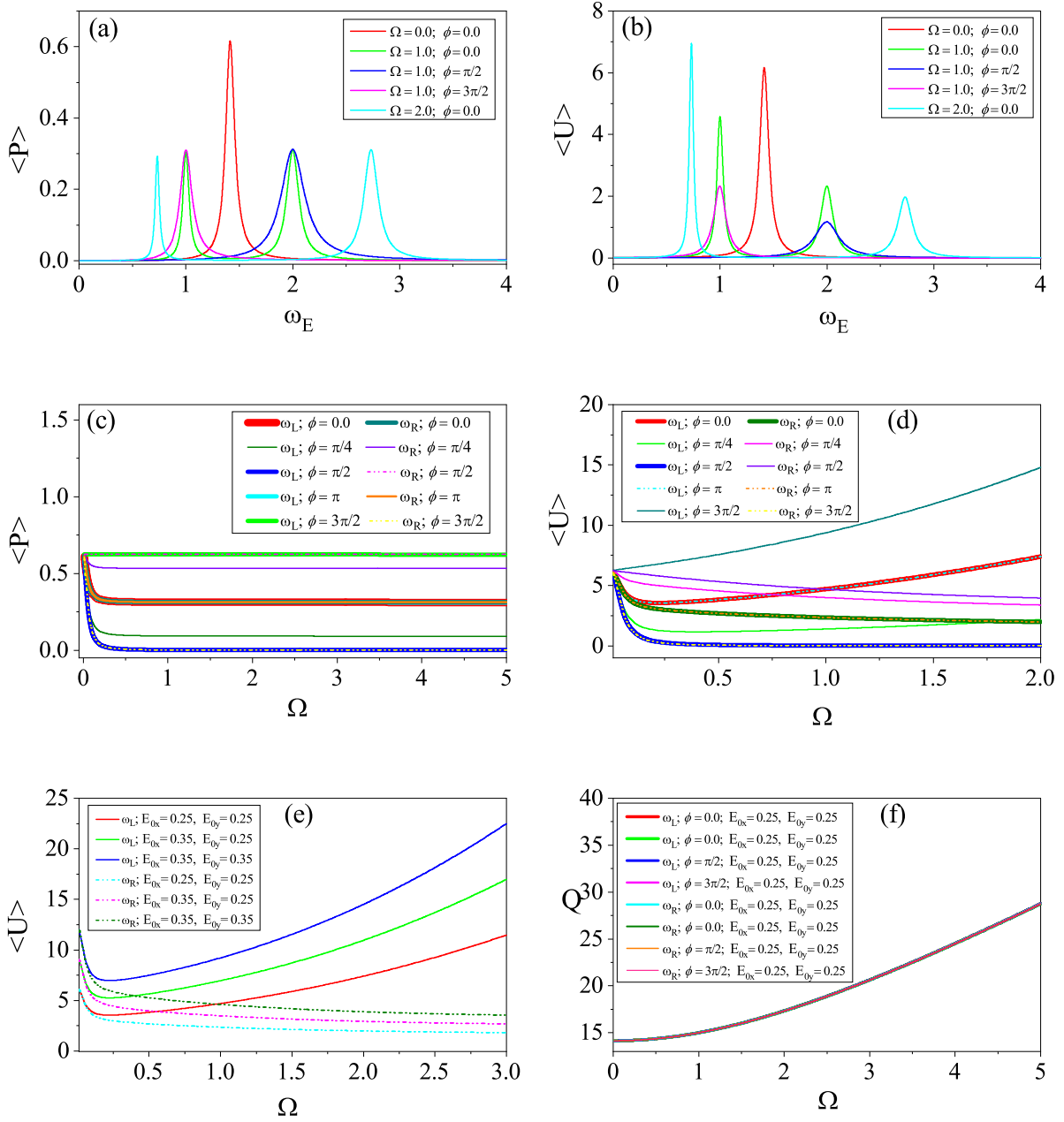}
\caption{{\footnotesize Demonstration about average power, average stored energy and their ratio.
Common parameter set for panels (a), (b), (c) and (d) :  $\omega^2 = 2.0, \gamma = 0.1$ and $E_{0x} = E_{0y} = 0.25$. Common parameter set for panels (e) and (f): $\omega^2 = 2.0$ and $\gamma = 0.1$.  (Units are arbitrary)}}
\label{fig.5}
\end{figure}

\noindent
In Fig.\ref{fig.5}, we have shown that how the average power and the mean stored energy depend on the driving frequency. Panel (a) shows that the asymmetric nature of the splitting of the spectrum of the average power is not so prominent as for the case of amplitude provided the phase difference between the two driving component is zero. It is a consequence of the following fact. The amplitude at the resonance condition  decreases with increase in resonating frequency as the frictional loss of energy becomes higher. The average power (\ref{en2}) depends on both these quantities as implied by Eqs. (\ref{eqr3}-\ref{eqr4}, \ref{en1}). Then the strongly asymmetric splitting of the amplitudes results the work done per unit time at lower resonating frequency with higher amplitude may be comparable to the same at higher frequency having lower amplitude. But the panel (b) implies that the spectrum of the stored  mean  energy may mimic the splitting pattern as that of the amplitudes. It may happen because the stored mean potential energy is not an explicit function of driving frequency (\ref{en3}) as like as the other part of the average energy.

Now one may be interested to compare the stored energy ($\langle U \rangle$) with the amount of work that is done in one cycle. Then the following relevant quantity \cite{feyn} of interest is given by

\begin{equation}\label{en4}
Q = 2 \pi \frac{\langle U \rangle}{\langle P \rangle 2 \pi/\omega_e} = \frac{\omega^2 + \omega_E^2}{2 \gamma \omega_E}   \;\;\;.
\end{equation}

\noindent
Thus $Q$ is defined as $2 \pi$ times the mean stored energy, divided by the work done per cycle. It may be an very useful quantity at resonance condition. In the presence of a magnetic field, for the resonance at lower frequency we represent $Q$ by $Q_L$. Other one is denoted by $Q_R$. Similarly one may represent $\langle P \rangle$ and $\langle U \rangle$ as $\langle P \rangle_L$ $\langle P \rangle_R$, $\langle U \rangle_L$ and $\langle U \rangle_R$, respectively. Panels (c), (d), (e) and (f) exhibit how these quantities depend on the different parameters. It is apparent here that at the resonance condition, the process of storing of the energy may be more efficient as the magnetic field becomes more stronger. It is also apparent here that although the stored energy depends on the amplitudes of the driving components and the phase difference between therm but the efficiency like quantity ($Q$) does not depend on these parameters. But the efficiency  depends on the parameters, frequency of the oscillator and damping strength, respectively. This aspect has been demonstrated in Fig.\ref{fig.5}.

\subsubsection{Driven damped cyclotron motion}

We now consider nature of the resonance phenomenon for the driven damped cyclotron motion. For $\omega = 0$, Eqs. (\ref{eqr15})-(\ref{eqr19}) become

\begin{equation}\label{eqr39}
H_0 = \left\{\omega_E^4 - \left(\Omega^2 - \gamma^2\right) \omega_E^2\right\}^2 + 4 \gamma^2 \Omega^2 \omega_E^4   \;\;\;,
\end{equation}

\begin{equation}\label{eqr40}
H_1 = \frac{-q\omega_E^2}{m} \left[\left\{\omega_E^4 - \left(\Omega^2 - \gamma^2\right) \omega_E^2\right\} E_{0x} + 2 \gamma \Omega \omega_E^2 E_{0y} \right]   \;\;\;,
\end{equation}

\begin{equation}\label{eqr41}
H_2 = \frac{q}{m} \omega_E \left[\left\{\omega_E^4 - \left(\Omega^2 - \gamma^2\right) \omega_E^2\right\} \left(\gamma E_{0x} - \Omega E_{0y}\right) + 2 \gamma \Omega \omega_E^2 \left(\gamma E_{0y} + \Omega E_{0x}\right)\right]   \;\;\;,
\end{equation}

\begin{equation}\label{eqr42}
H_3 =  \frac{-q\omega_E^2}{m} \left[\left\{ \omega_E^4 - \left(\Omega^2 - \gamma^2\right) \omega_E^2\right\} E_{0y} - 2 \gamma \Omega \omega_E^2 E_{0x} \right]   \;\;\;,
\end{equation}

\noindent
and

\begin{equation}\label{eqr43}
H_4 = \frac{q}{m} \omega_E \left[\left\{ \omega_E^4 - \left(\Omega^2 - \gamma^2\right) \omega_E^2\right\} \left(\gamma E_{0y} + \Omega E_{0x}\right) - 2 \gamma \Omega \omega_E^2 \left(\gamma E_{0x} - \Omega E_{0y}\right)\right]   \;\;\;.
\end{equation}

\noindent
Making use of the above relations in Eqs. (\ref{eqr11}-\ref{eqr12}) we find again that it is difficult to determine the resonance condition maximizing $a$ and $b$ with respect to the driving frequency. Then one may adopt the previous trick. At the weak damping limit when the resonance phenomenon may be appear then $H_0$ in Eq. (\ref{eqr39}) may be minimum around the following condition,

\begin{equation}\label{eqr44}
\omega_E^4 - \Omega^2 \omega_E^2 = 0  \;\;\;,
\end{equation}

\noindent
Thus the approximate resonance condition may be read as

\begin{equation}\label{eqr45}
\omega_E \simeq \Omega \;\;\;.
\end{equation}

\begin{figure}[!htb]
\includegraphics[width=16cm,angle=0,clip]{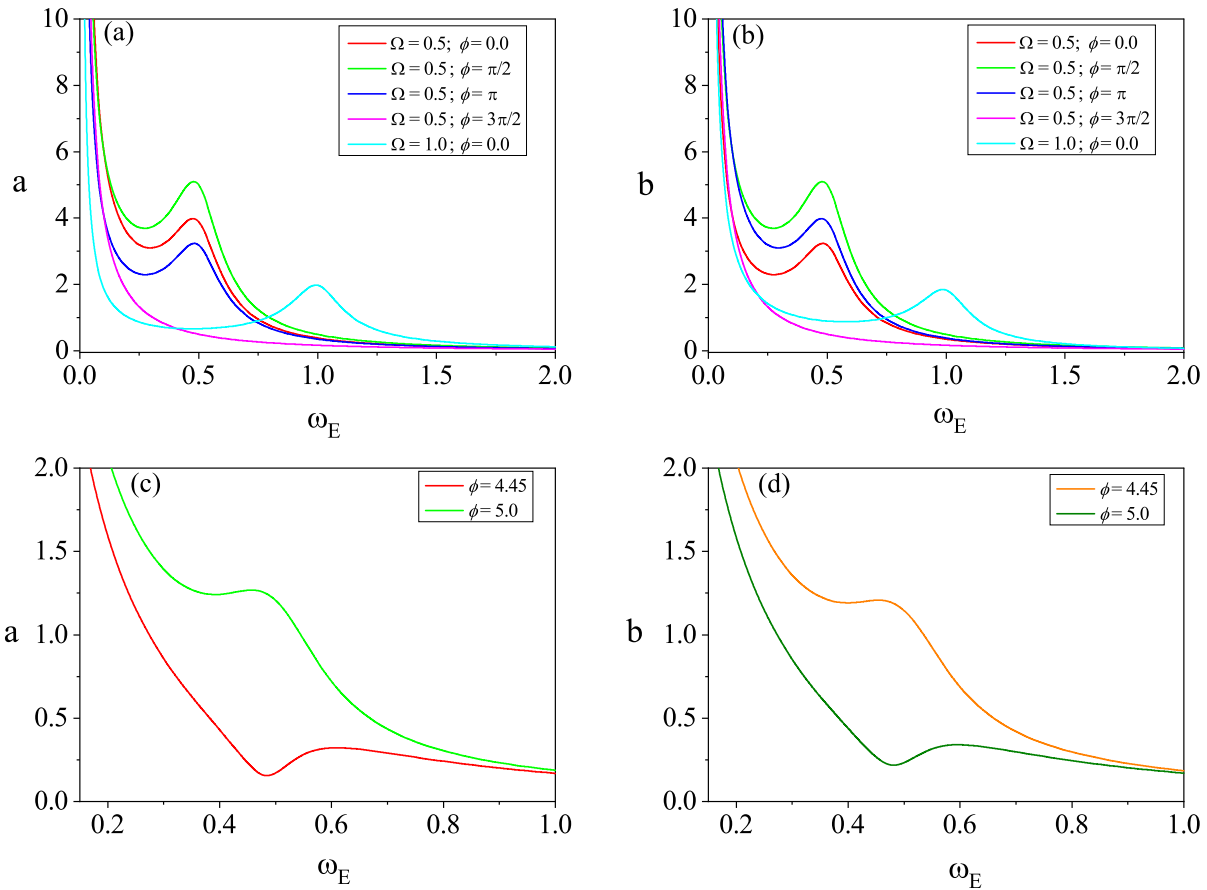}
\caption{{\footnotesize Demonstration about amplitudes and anti resonance for the driven damped oscillator. (a) and (b) Plot of $a$ and $b$ vs. $\omega_E$ for different values of $\Omega$ and $\phi$ along with the paramete set: $\omega^2 = 0.0, \gamma = 0.1$ and $ E_{0x} = E_{0y} = 0.25$; (c) and (d) Plot of $a$ and $b$ vs. $\omega_E$ for different values of $\phi$ along with the paramete set: $\omega^2 = 0.0, \gamma = 0.1$ $\Omega = 0.5$ and $ E_{0x} = E_{0y} = 0.25$ (Units are arbitrary).}}
\label{fig.6}
\end{figure}

\noindent
Another root as a solution of Eq. (\ref{eqr45}), $\omega_E = 0$. It corresponds to the diverging motion of the charged particle in the presence of a constant electromagnetic field with $ {\bf E} = \hat{i} E_{0x} + \hat{j} E_{0y}$. One may verify it easily considering the relevant equations of motion. Thus we may have two peaks at $\omega_E=0$ and $\omega_E = \Omega$, respectively. This is implied by  Eqs.(\ref{eq5}-\ref{eq6}). To check accuracy of the approximation we have demonstrated the exact results in Fig.\ref{fig.6}. The resonance conditions according to this figure compared with the analytically calculated results in Table \ref{tab.3}. It shows that there is a very good agreement between the approximate and the exact results.

\begin{table}[ht]
\caption{Comparison between theoretically calculated resonating frequency and the exact result for the driven damped cyclotron motion}
\begin{center}
\begin{tabular}{|c|c|c|c|c|}
\hline
Value of $\Omega$ & Theoretical & Exact \\ 
\hline
0.5 & 0.50 & 0.48 \\
\hline
1.0 & 1.00 & 0.99 \\
\hline
1.5 & 1.50 & 1.49 \\
\hline
2.0 & 2.00 & 1.99 \\
\hline
\end{tabular}
\end{center}
\label{tab.3}
\end{table}

The remaining panels of Fig.\ref{fig.6} imply that the anti resonance phenomenon may appear even for a system with only non conservative force fields. For this system both resonance and anti resonance phenomena may appear at the same driving frequency depending upon the phase difference between the input signals.  Thus the underlying reason to appear the anti resonance for the driven damped cyclotron motion may be the same as that of driven damped harmonic oscillator in the presence of a magnetic field.

\noindent
Before leaving this part we would mention here that making use of relations (\ref{eqr11}-\ref{eqr12}) with Eqs. (\ref{eqr39}-\ref{eqr43}) into the previous subsection one may study the energetics for the driven damped cyclotron motion.

\section{Steady state dynamics of a three dimensional harmonic oscillator in the presence of an electromagnetic field: Similarity of the output spectrum to the Normal Zeeman effect}

We now consider the steady state dynamics for the general case, driven damped three dimensional harmonic oscillator in the presence of a magnetic field (${\bf B} = (B_x, B_y, B_z)$) which may be at arbitrary direction. For this case we choose a periodic electric field as

\begin{equation}\label{anis1}
{\bf E} = \hat{i} E_{0x} \cos(\omega_E t - \phi_x) + \hat{j} E_{0y} \cos(\omega_E t - \phi_y) + \hat{k} E_{0z} \cos(\omega_E t - \phi_z)   \;\;\;
\end{equation}

\noindent
where $\phi_x$, $\phi_y$ and $\phi_z$ are the phase constants associated with each component along respective directions. Then one may write the equations of motion as

\begin{equation}\label{anis2}
m \ddot{x} = - m \omega_x^2 x - m \gamma \dot{x} + m \Omega_z \dot{y} - m \Omega_y \dot{z}+ q E_{0x} \cos(\omega_E t - \phi_x)   \;\;\;,
\end{equation}

\begin{equation}\label{anis3}
m \ddot{y} = - m \omega_y^2 y - m \gamma \dot{y} + m \Omega_x \dot{z} - m \Omega_z \dot{x}+ q E_{0y} \cos(\omega_E t - \phi_y)   \;\;\;,
\end{equation}

\begin{equation}\label{anis4}
m \ddot{z} = - m \omega_z^2 z - m \gamma \dot{z} + m \Omega_y \dot{x} - m \Omega_x \dot{y}+ q E_{0z} \cos(\omega_E t - \phi_z)   \;\;\;.
\end{equation}

\noindent
Here we have used $\Omega_x = q B_x/m, \Omega_y = q B_y/m$ and $\Omega_x = q B_z/m$. Increase in dimension of the system makes difficult to study the dynamics at short time in the absence of electric field. However, to characterize the steady state dynamics we choose the following particular solutions to solve the above equations

\begin{equation}\label{anis5}
x(t) = a \cos(\omega_E t - \phi_1)   \;\;\;,
\end{equation}

\begin{equation}\label{anis6}
y(t) = b \cos(\omega_E t - \phi_2)   \;\;\; ,
\end{equation}

\noindent
and

\begin{equation}\label{anis7}
z(t) = c \cos(\omega_E t - \phi_3)   \;\;\;.
\end{equation}

\noindent
Following the earlier case now one may determine the amplitudes and the phase constants as

\begin{equation}\label{anis8}
a =  \frac{\sqrt{A_1^2 + A_2^2}}{D_1}  \;\;\;,
\end{equation}

\begin{equation}\label{anis9}
b = \frac{\sqrt{B_1^2 + B_2^2}}{D_1}   \;\;\;,
\end{equation}

\begin{equation}\label{anis10}
c = \frac{\sqrt{C_1^2 + C_2^2}}{D_2}   \;\;\;.
\end{equation}

\begin{equation}\label{anis11}
\tan \phi_1 = \frac{A_2}{A_1}   \;\;\;,
\end{equation}

\begin{equation}\label{anis12}
\tan \phi_2 = \frac{B_2}{B_1}   \;\;\;,
\end{equation}

\noindent
and

\begin{equation}\label{anis13}
\tan \phi_3 = \frac{C_2}{C_1}   \;\;\;.
\end{equation}

\noindent
In these relations we have used

\begin{equation}\label{anis14}
A_1 = \left(H_c H_x - H_z H_2\right) \left(H_b H_c + H_3 H_6\right) + \left(H_c H_y + H_z H_3\right) \left(H_c H_1 + H_2 H_6\right)   \;\;\;,
\end{equation}

\begin{equation}\label{anis15}
A_2 = \left(H_c H_x^\prime - H_z^\prime H_2\right) \left(H_b H_c + H_3 H_6\right) + \left(H_c H_y^\prime + H_z^\prime H_3\right) \left(H_c H_1 + H_2 H_6\right)   \;\;\;,
\end{equation}

\begin{equation}\label{anis16}
B_1 = \left(H_c H_y + H_z H_3\right) \left(H_a H_c + H_2 H_5\right) - \left(H_c H_x - H_z H_2\right) \left(H_c H_4 - H_3 H_5\right)   \;\;\;,
\end{equation}

\begin{equation}\label{anis17}
B_2 = \left(H_c H_y^\prime + H_z^\prime H_3\right) \left(H_a H_c + H_2 H_5\right) - \left(H_c H_x^\prime - H_z^\prime H_2\right) \left(H_c H_4 - H_3 H_5\right)   \;\;\;,
\end{equation}

\begin{equation}\label{anis18}
C_1 = \left(H_b H_z - H_y H_6\right) \left(H_a H_b + H_1 H_4\right) + \left(H_b H_x + H_y H_1\right) \left(H_b H_5 + H_4 H_6\right)   \;\;\;,
\end{equation}

\begin{equation}\label{anis19}
C_2 = \left(H_b H_z^\prime - H_y^\prime H_6\right) \left(H_a H_b + H_1 H_4\right) + \left(H_b H_x^\prime + H_y H_1\right) \left(H_b H_5 + H_4 H_6\right)   \;\;\;,
\end{equation}

\begin{equation}\label{anis20}
D_1 = \left(H_a H_c + H_2 H_5\right) \left(H_b H_c + H_3 H_6\right) + \left(H_c H_1 + H_2 H_6\right) \left(H_c H_4 - H_3 H_5\right)   \;\;\;.
\end{equation}

\begin{equation}\label{anis21}
D_2 = \left(H_a H_b + H_1 H_4\right) \left(H_b H_c + H_3 H_6\right) + \left(H_b H_5 + H_4 H_6\right) \left(H_b H_2 - H_1 H_3\right)   \;\;\;,
\end{equation}

\noindent
with

\begin{eqnarray}\label{anis22}
H_a &=& \left(\omega_x^2 - \omega_E^2\right)^2 \left(\omega_y^2 - \omega_E^2\right) \left(\omega_z^2 - \omega_E^2\right) + \Big\{\gamma^2 \left(\omega_y^2 - \omega_E^2\right) \left(\omega_z^2 - \omega_E^2\right)   \nonumber \\ \nonumber \\
&-& \Omega_y^2 \left(\omega_x^2 - \omega_E^2\right) \left(\omega_y^2 - \omega_E^2\right) - \Omega_z^2 \left(\omega_z^2 - \omega_E^2\right) \left(\omega_x^2 - \omega_E^2\right)\Big\} \omega_E^2   \;\;\;,
\end{eqnarray}

\begin{eqnarray}\label{anis23}
H_b &=& \left(\omega_x^2 - \omega_E^2\right) \left(\omega_y^2 - \omega_E^2\right)^2 \left(\omega_z^2 - \omega_E^2\right) + \Big\{\gamma^2 \left(\omega_z^2 - \omega_E^2\right) \left(\omega_x^2 - \omega_E^2\right)   \nonumber \\ \nonumber \\
&-& \Omega_z^2 \left(\omega_y^2 - \omega_E^2\right) \left(\omega_z^2 - \omega_E^2\right) - \Omega_x^2 \left(\omega_x^2 - \omega_E^2\right) \left(\omega_y^2 - \omega_E^2\right)\Big\} \omega_E^2   \;\;\;,
\end{eqnarray}

\begin{eqnarray}\label{anis24}
H_c &=& \left(\omega_x^2 - \omega_E^2\right) \left(\omega_y^2 - \omega_E^2\right) \left(\omega_z^2 - \omega_E^2\right)^2 + \Big\{\gamma^2 \left(\omega_x^2 - \omega_E^2\right) \left(\omega_y^2 - \omega_E^2\right)   \nonumber \\ \nonumber \\
&-& \Omega_x^2 \left(\omega_z^2 - \omega_E^2\right) \left(\omega_x^2 - \omega_E^2\right) - \Omega_y^2 \left(\omega_y^2 - \omega_E^2\right) \left(\omega_z^2 - \omega_E^2\right)\Big\} \omega_E^2   \;\;\;,
\end{eqnarray}

\begin{eqnarray}\label{anis25}
H_x &=& \frac{q}{m} \Big\{E_{0x} \left(\omega_x^2 - \omega_E^2\right) \left(\omega_y^2 - \omega_E^2\right) \left(\omega_z^2 - \omega_E^2\right) \cos \phi_x   \nonumber \\ \nonumber \\
&-& E_{0x} \gamma \omega_E \left(\omega_y^2 - \omega_E^2\right) \left(\omega_z^2 - \omega_E^2\right) \sin \phi_x   \nonumber \\ \nonumber \\
&+& E_{0y} \Omega_z \omega_E \left(\omega_z^2 - \omega_E^2\right) \left(\omega_x^2 - \omega_E^2\right) \sin \phi_y   \nonumber \\ \nonumber \\
&-& E_{0z} \Omega_y \omega_E \left(\omega_x^2 - \omega_E^2\right) \left(\omega_y^2 - \omega_E^2\right) \sin \phi_z\Big\}   \;\;\;,
\end{eqnarray}

\begin{eqnarray}\label{anis25}
H_y &=& \frac{q}{m} \Big\{E_{0y} \left(\omega_x^2 - \omega_E^2\right) \left(\omega_y^2 - \omega_E^2\right) \left(\omega_z^2 - \omega_E^2\right) \cos \phi_y   \nonumber \\ \nonumber \\
&-& E_{0x} \Omega_z \omega_E \left(\omega_y^2 - \omega_E^2\right) \left(\omega_z^2 - \omega_E^2\right) \sin \phi_x   \nonumber \\ \nonumber \\
&-& E_{0y} \gamma \omega_E \left(\omega_z^2 - \omega_E^2\right) \left(\omega_x^2 - \omega_E^2\right) \sin \phi_y   \nonumber \\ \nonumber \\
&+& E_{0z} \Omega_x \omega_E \left(\omega_x^2 - \omega_E^2\right) \left(\omega_y^2 - \omega_E^2\right) \sin \phi_z\Big\}   \;\;\;,
\end{eqnarray}

\begin{eqnarray}\label{anis25}
H_z &=& \frac{q}{m} \Big\{E_{0z} \left(\omega_x^2 - \omega_E^2\right) \left(\omega_y^2 - \omega_E^2\right) \left(\omega_z^2 - \omega_E^2\right) \cos \phi_z   \nonumber \\ \nonumber \\
&+& E_{0x} \Omega_y \omega_E \left(\omega_y^2 - \omega_E^2\right) \left(\omega_z^2 - \omega_E^2\right) \sin \phi_x   \nonumber \\ \nonumber \\
&-& E_{0y} \Omega_x \omega_E \left(\omega_z^2 - \omega_E^2\right) \left(\omega_x^2 - \omega_E^2\right) \sin \phi_y   \nonumber \\ \nonumber \\
&-& E_{0z} \gamma \omega_E \left(\omega_x^2 - \omega_E^2\right) \left(\omega_y^2 - \omega_E^2\right) \sin \phi_z\Big\}   \;\;\;,
\end{eqnarray}

\begin{eqnarray}\label{anis26}
H_x^\prime &=& \frac{q}{m} \Big\{E_{0x} \left(\omega_x^2 - \omega_E^2\right) \left(\omega_y^2 - \omega_E^2\right) \left(\omega_z^2 - \omega_E^2\right) \sin \phi_x   \nonumber \\ \nonumber \\
&+& E_{0x} \gamma \omega_E \left(\omega_y^2 - \omega_E^2\right) \left(\omega_z^2 - \omega_E^2\right) \cos \phi_x   \nonumber \\ \nonumber \\
&-& E_{0y} \Omega_z \omega_E \left(\omega_z^2 - \omega_E^2\right) \left(\omega_x^2 - \omega_E^2\right) \cos \phi_y   \nonumber \\ \nonumber \\
&+& E_{0z} \Omega_y \omega_E \left(\omega_x^2 - \omega_E^2\right) \left(\omega_y^2 - \omega_E^2\right) \cos \phi_z\Big\}   \;\;\;,
\end{eqnarray}

\begin{eqnarray}\label{anis27}
H_y^\prime &=& \frac{q}{m} \Big\{E_{0y} \left(\omega_x^2 - \omega_E^2\right) \left(\omega_y^2 - \omega_E^2\right) \left(\omega_z^2 - \omega_E^2\right) \sin \phi_y   \nonumber \\ \nonumber \\
&+& E_{0x} \Omega_z \omega_E \left(\omega_y^2 - \omega_E^2\right) \left(\omega_z^2 - \omega_E^2\right) \cos \phi_x   \nonumber \\ \nonumber \\
&+& E_{0y} \gamma \omega_E \left(\omega_z^2 - \omega_E^2\right) \left(\omega_x^2 - \omega_E^2\right) \cos \phi_y   \nonumber \\ \nonumber \\
&-& E_{0z} \Omega_x \omega_E \left(\omega_x^2 - \omega_E^2\right) \left(\omega_y^2 - \omega_E^2\right) \cos \phi_z\Big\}   \;\;\;,
\end{eqnarray}

\begin{eqnarray}\label{anis28}
H_z^\prime &=& \frac{q}{m} \Big\{E_{0z} \left(\omega_x^2 - \omega_E^2\right) \left(\omega_y^2 - \omega_E^2\right) \left(\omega_z^2 - \omega_E^2\right) \sin \phi_z   \nonumber \\ \nonumber \\
&-& E_{0x} \Omega_y \omega_E \left(\omega_y^2 - \omega_E^2\right) \left(\omega_z^2 - \omega_E^2\right) \cos \phi_x   \nonumber \\ \nonumber \\
&+& E_{0y} \Omega_x \omega_E \left(\omega_z^2 - \omega_E^2\right) \left(\omega_x^2 - \omega_E^2\right) \cos \phi_y   \nonumber \\ \nonumber \\
&+& E_{0z} \gamma \omega_E \left(\omega_x^2 - \omega_E^2\right) \left(\omega_y^2 - \omega_E^2\right) \cos \phi_z\Big\}   \;\;\;,
\end{eqnarray}

\begin{equation}\label{anis29}
H_1 = \Big\{\gamma \Omega_z \left(\omega_y^2 - \omega_E^2\right) \left(\omega_z^2 - \omega_E^2\right) + \gamma \Omega_z \left(\omega_z^2 - \omega_E^2\right) \left(\omega_x^2 - \omega_E^2\right) - \Omega_x \Omega_y \left(\omega_x^2 - \omega_E^2\right) \left(\omega_y^2 - \omega_E^2\right)\Big\} \omega_E^2   \;\;\;,
\end{equation}

\begin{equation}\label{anis30}
H_2 = \Big\{\gamma \Omega_y \left(\omega_x^2 - \omega_E^2\right) \left(\omega_y^2 - \omega_E^2\right) + \gamma \Omega_y \left(\omega_y^2 - \omega_E^2\right) \left(\omega_z^2 - \omega_E^2\right) + \Omega_z \Omega_x \left(\omega_z^2 - \omega_E^2\right) \left(\omega_x^2 - \omega_E^2\right)\Big\} \omega_E^2   \;\;\;,
\end{equation}

\begin{equation}\label{anis31}
H_3 = \Big\{\gamma \Omega_x \left(\omega_x^2 - \omega_E^2\right) \left(\omega_y^2 - \omega_E^2\right) + \gamma \Omega_x \left(\omega_z^2 - \omega_E^2\right) \left(\omega_x^2 - \omega_E^2\right) - \Omega_y \Omega_z \left(\omega_y^2 - \omega_E^2\right) \left(\omega_z^2 - \omega_E^2\right)\Big\} \omega_E^2   \;\;\;,
\end{equation}

\begin{equation}\label{anis32}
H_4 = \Big\{\gamma \Omega_z \left(\omega_y^2 - \omega_E^2\right) \left(\omega_z^2 - \omega_E^2\right) + \gamma \Omega_z \left(\omega_z^2 - \omega_E^2\right) \left(\omega_x^2 - \omega_E^2\right) + \Omega_x \Omega_y \left(\omega_x^2 - \omega_E^2\right) \left(\omega_y^2 - \omega_E^2\right)\Big\} \omega_E^2   \;\;\;,
\end{equation}

\begin{equation}\label{anis33}
H_5 = \Big\{\gamma \Omega_y \left(\omega_x^2 - \omega_E^2\right) \left(\omega_y^2 - \omega_E^2\right) + \gamma \Omega_y \left(\omega_y^2 - \omega_E^2\right) \left(\omega_z^2 - \omega_E^2\right) - \Omega_z \Omega_x \left(\omega_z^2 - \omega_E^2\right) \left(\omega_x^2 - \omega_E^2\right)\Big\} \omega_E^2   \;\;\;,
\end{equation}

\noindent
and

\begin{equation}\label{anis34}
H_6 = \Big\{\gamma \Omega_x \left(\omega_x^2 - \omega_E^2\right) \left(\omega_y^2 - \omega_E^2\right) + \gamma \Omega_x \left(\omega_z^2 - \omega_E^2\right) \left(\omega_x^2 - \omega_E^2\right) + \Omega_y \Omega_z \left(\omega_y^2 - \omega_E^2\right) \left(\omega_z^2 - \omega_E^2\right)\Big\} \omega_E^2   \;\;\;.
\end{equation}

\noindent
One may now show easily that Eqs. (\ref{anis8})-(\ref{anis13}) reduce to the results of the previous section. It constitutes an important check of our calculation.

\begin{figure}[!htb]
\includegraphics[width=16cm,angle=0,clip]{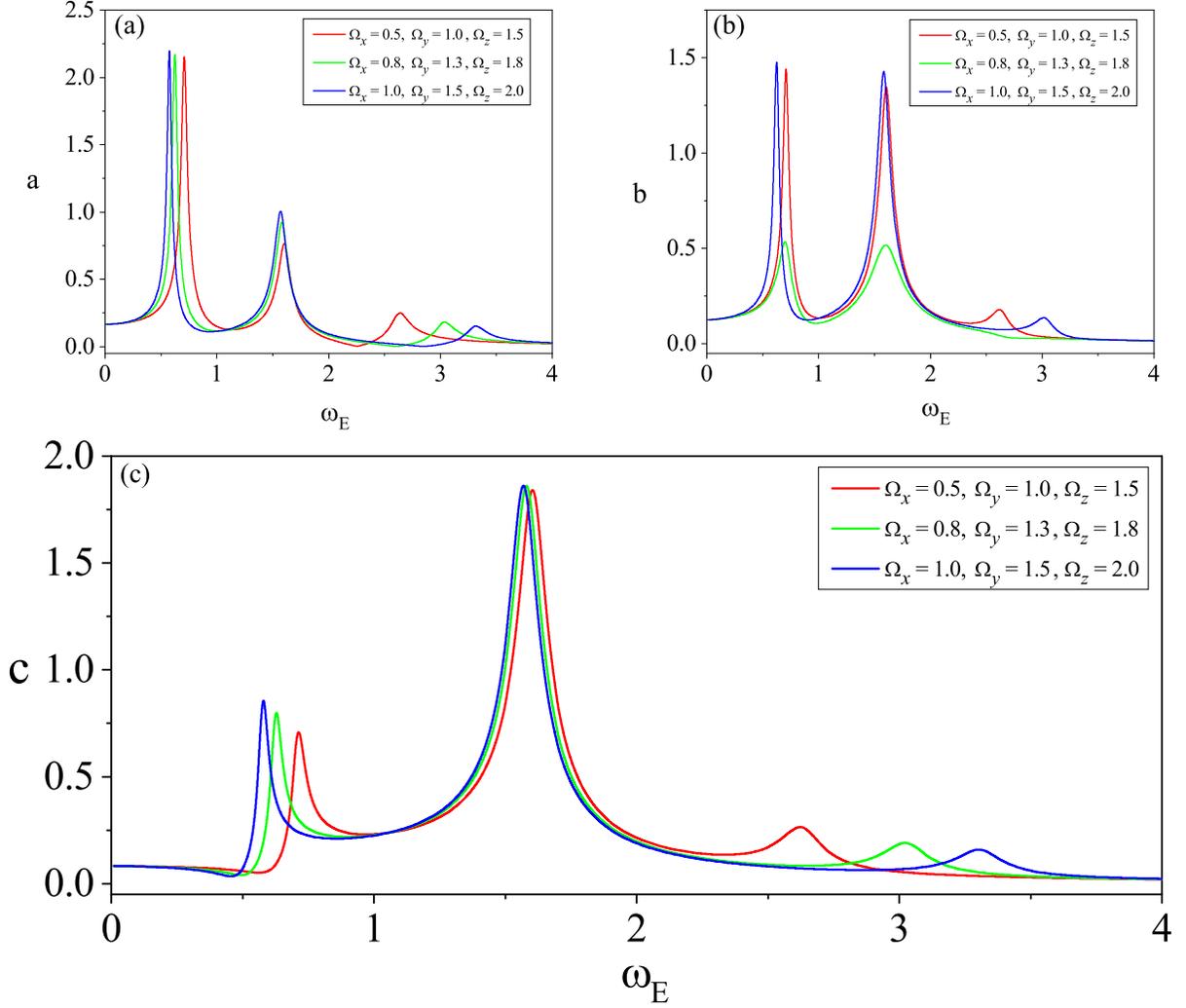}
\caption{{\footnotesize Plot of $a$, $b$ and $c$ vs. $\omega_E$ for different values of $\Omega_x$, $\Omega_y$ and $\Omega_z$ alon with the parameter set: $\omega_x^2 = 1.5, \omega_y^2 = 2.0, \omega_z^2 = 3.0, \gamma = 0.1, E_{0x} = E_{0y} = E_{0z} = 0.25$ and $\phi_x = \phi_y = \phi_z = 0.0$ (Units are arbitrary).}}
\label{fig.7}
\end{figure}

\noindent
We are now in a position to determine the resonance condition. The nature of the amplitude functions implies that it is very difficult to have the condition even applying the previous technique. According to the earlier procedure, we have to solve the following equation,

{\footnotesize
\begin{eqnarray}\label{anis35}
&& \left(\omega_x^2 - \omega_E^2\right)^2 \left(\omega_y^2 - \omega_E^2\right)^2 \left(\omega_z^2 - \omega_E^2\right)^2 \bigg[\Big\{\left(\omega_x^2 - \omega_E^2\right)^2 \left(\omega_y^2 - \omega_E^2\right) \left(\omega_z^2 - \omega_E^2\right)^2 - \Omega_x^2 \left(\omega_x^2 - \omega_E^2\right)^2 \left(\omega_z^2 - \omega_E^2\right) \omega_E^2   \nonumber \\
&-& 2 \Omega_y^2 \left(\omega_x^2 - \omega_E^2\right) \left(\omega_y^2 - \omega_E^2\right) \left(\omega_z^2 - \omega_E^2\right)\omega_E^2 - \Omega_z^2 \left(\omega_x^2 - \omega_E^2\right) \left(\omega_z^2 - \omega_E^2\right)^2 \omega_E^2 + \Omega_x^2 \Omega_y^2 \left(\omega_x^2 - \omega_E^2\right) \omega_E^4   \nonumber \\
&+& \Omega_y^4 \left(\omega_y^2 - \omega_E^2\right) \omega_E^4 + \Omega_y^2 \Omega_z^2 \left(\omega_z^2 - \omega_E^2\right) \omega_E^4\Big\} \times \Big\{\left(\omega_x^2 - \omega_E^2\right) \left(\omega_y^2 - \omega_E^2\right)^2 \left(\omega_z^2 - \omega_E^2\right)^2   \nonumber \\
&-& 2 \Omega_x^2 \left(\omega_x^2 - \omega_E^2\right) \left(\omega_y^2 - \omega_E^2\right) \left(\omega_z^2 - \omega_E^2\right) \omega_E^2 - \Omega_y^2 \left(\omega_y^2 - \omega_E^2\right)^2 \left(\omega_z^2 - \omega_E^2\right) \omega_E^2 - \Omega_z^2 \left(\omega_y^2 - \omega_E^2\right) \left(\omega_z^2 - \omega_E^2\right)^2 \omega_E^2   \nonumber \\
&+& \Omega_x^4 \left(\omega_x^2 - \omega_E^2\right) \omega_E^4 + \Omega^2 \Omega_y^2 \left(\omega_y^2 - \omega_E^2\right) \omega_E^4 + \Omega_z^2 \Omega_x^2 \left(\omega_z^2 - \omega_E^2\right) \omega_E^4\Big\} - \Omega_x^2 \Omega_y^2 \Big\{\Omega_x^2 \left(\omega_x^2 - \omega_E^2\right) \omega_E^2   \nonumber \\
&+& \Omega_y^2 \left(\omega_y^2 - \omega_E^2\right) \omega_E^2 + \Omega_z^2 \left(\omega_z^2 - \omega_E^2\right) \omega_E^2 - \left(\omega_x^2 - \omega_E^2\right)^2 \left(\omega_y^2 - \omega_E^2\right)^2 \left(\omega_z^2 - \omega_E^2\right)^2\Big\}^2 \omega_E^4\bigg] = 0  \;\;\;,
\end{eqnarray}
}

\noindent
to find the values of $\omega_E$ for which $a$ and $b$ may be maximum. Similarly, one may face difficulty to fine the  conditions for which $c$ may be maximum. Then we may be interested to find the resonance condition for relatively simple case like $\omega_x = \omega_y = \omega_z = \omega$, $\Omega_x \neq \Omega_y \neq \Omega_z \neq \Omega$. For this case to determine the conditions for which $a$ and $b$ may be maximum we have to solve the following equation,

{\footnotesize
\begin{eqnarray}\label{anis36}
&& \omega_E^{16} - \left(a_1 + a_2 + 8 \omega^2\right) \omega_E^{14} + \left\{28 \omega^4 + 6 \left(a_1 + a_2\right) \omega^2 + \left(a_1 a_2 - a_2 a_1 + b_1 + b_2\right)\right\} \omega_E^{12}   \nonumber \\
&-& \left\{56 \omega^6 + 15 \left(a_1 + a_2\right) \omega^4 + 4 \left(a_1 a_2 - a_2 a_1 + b_1 + b_2\right) \omega^2 - \left(a_1 b_2 + a_2 b_1 + 2 a_3 b_3\right)\right\} \omega_E^{10}   \nonumber \\
&+& \left\{70 \omega^8 + 20 \left(a_1 + a_2\right) \omega^6 + 6 \left(a_1 a_2 - a_2 a_1 + b_1 + b_2\right) \omega^4 + 2 \left(a_1 b_2 + a_2 b_1 + 2 a_3 b_3\right) \omega^2 + \left(b_1 b_2 - b_3^2\right)\right\} \omega_E^{8}   \nonumber \\
&-& \left\{56 \omega^6 + 15 \left(a_1 + a_2\right) \omega^4 + 4 \left(a_1 a_2 - a_2 a_1 + b_1 + b_2\right) \omega^2 - \left(a_1 b_2 + a_2 b_1 + 2 a_3 b_3\right)\right\} \omega^4 \omega_E^{6}   \nonumber \\
&-& \left\{28 \omega^4 + 6 \left(a_1 + a_2\right) \omega^2 + \left(a_1 a_2 - a_2 a_1 + b_1 + b_2\right)\right\} \omega^8 \omega_E^{4} - \left(a_1 + a_2 + 8 \omega^2\right) \omega^{12} \omega_E^{2} + \omega^{16} =0  \;\;\;
\end{eqnarray}
}

\noindent
where $a_1 = \Omega_x^2 + 2 \Omega_y^2 + \Omega_z^2$, 
$a_2 = 2 \Omega_x^2 + \Omega_y^2 + \Omega_z^2$, 
$a_3 = \Omega_x \Omega_y$,  
$b_1 = \Omega_y^2 \left(\Omega_x^2 + \Omega_y^2 + \Omega_z^2\right)$,  
$b_2 = \Omega_x^2 \left(\Omega_x^2 + \Omega_y^2 + \Omega_z^2\right)$ and 
$b_3 = \Omega_x \Omega_y \left(\Omega_x^2 + \Omega_y^2 + \Omega_z^2\right)$. It seems to be  difficult to solve the above equation. Similarly, one may face difficulty to find the conditions for which $c$ may be maximum. Then we may determine the resonance condition from the plot as shown in the panels (a), (b) and (c) in Fig.\ref{fig.7}. It exhibits an interesting feature that in the process of asymmetric splitting one more additional peak may appear around the frequency of the relevant vibrational mode. To understand the appearance of the three peaks we consider the simplest case, $\omega_x = \omega_y = \omega_z = \omega$, $\Omega_x = \Omega_y = \Omega_z = \Omega$. For this condition, Eqs. (\ref{anis14})-(\ref{anis21}) become

\begin{equation}\label{anis37}
A_1 = \left(H_x H_0 - H_z H_2\right) \left(H_0^2 + H_1 H_2\right) + \left(H_y H_0 + H_z H_1\right) \left(H_2^2 + H_0 H_1\right)   \;\;\;,
\end{equation}

\begin{equation}\label{anis38}
A_2 = \left(H_x^\prime H_0 - H_z^\prime H_2\right) \left(H_0^2 + H_1 H_2\right) + \left(H_y^\prime H_0 + H_z^\prime H_1\right) \left(H_2^2 + H_0 H_1\right)   \;\;\;,
\end{equation}

\begin{equation}\label{anis40}
B_1 = \left(H_y H_0 + H_z H_1\right) \left(H_0^2 + H_1 H_2\right) + \left(H_x H_0 - H_z H_2\right) \left(H_1^2 - H_0 H_2\right)   \;\;\;,
\end{equation}

\begin{equation}\label{anis41}
B_2 = \left(H_y^\prime H_0 + H_z^\prime H_1\right) \left(H_0^2 + H_1 H_2\right) + \left(H_x^\prime H_0 - H_z^\prime H_2\right) \left(H_1^2 - H_0 H_2\right)   \;\;\;,
\end{equation}

\begin{equation}\label{anis42}
C_1 = \left(H_z H_0 + H_x H_1\right) \left(H_0^2 + H_1 H_2\right) + \left(H_y H_0 - H_x H_2\right) \left(H_1^2 - H_0 H_2\right)   \;\;\;,
\end{equation}

\begin{equation}\label{anis43}
C_2 = \left(H_z^\prime H_0 + H_x^\prime H_1\right) \left(H_0^2 + H_1 H_2\right) + \left(H_y^\prime H_0 - H_x^\prime H_2\right) \left(H_1^2 - H_0 H_2\right)   \;\;\;,
\end{equation}

\begin{equation}\label{anis44}
D_1 = D_2 = D = \left(H_0^2 + H_1 H_2\right)^2 - \left(H_1^2 - H_0 H_2\right) \left(H_2^2 + H_0 H_1\right)   \;\;\;,
\end{equation}

\noindent
with

\begin{equation}\label{anis45}
H_0 = \left(\omega^2 - \omega_E^2\right)^2 - \left(2 \Omega^2 - \gamma^2\right) \omega_E^2   \;\;\;,
\end{equation}

\begin{equation}\label{anis46}
H_1 = \Omega \omega_E^2 \left(2 \gamma - \Omega\right)   \;\;\;,
\end{equation}

\begin{equation}\label{anis47}
H_2 = \Omega \omega_E^2 \left(2 \gamma + \Omega\right)   \;\;\;,
\end{equation}

\begin{equation}\label{anis48}
H_x = \frac{q}{m} E_{0x} \left(\omega^2 - \omega_E^2\right)   \;\;\;,
\end{equation}

\begin{equation}\label{anis49}
H_y = \frac{q}{m} E_{0y} \left(\omega^2 - \omega_E^2\right)   \;\;\;,
\end{equation}

\begin{equation}\label{anis50}
H_z = \frac{q}{m} E_{0z} \left(\omega^2 - \omega_E^2\right)   \;\;\;,
\end{equation}

\begin{equation}\label{anis51}
H_x^\prime = \frac{q}{m} \omega_E \left(\gamma E_{0x} - \Omega E_{0y} + \Omega E_{0z}\right)   \;\;\;,
\end{equation}

\begin{equation}\label{anis52}
H_y^\prime = \frac{q}{m} \omega_E \left(\gamma E_{0y} - \Omega E_{0z} + \Omega E_{0x}\right)   \;\;\;,
\end{equation}

\noindent
and

\begin{equation}\label{anis53}
H_z^\prime = \frac{q}{m} \omega_E \left(\gamma E_{0z} - \Omega E_{0x} + \Omega E_{0y}\right)   \;\;\;.
\end{equation}

\noindent
To determine the resonance condition we now follow the previous section. At the weak damping limit when the resonance phenomenon may appear then $D$ may be minimum around the following condition,

\begin{equation}\label{anis54}
\left(H_0^2 + H_1 H_2\right)^2 - \left(H_1^2 - H_0 H_2\right) \left(H_2^2 + H_0 H_1\right) = 0   \;\;\;.
\end{equation}

\noindent
where

\begin{equation}\label{anis55}
H_0 = \left(\omega^2 - \omega_E^2\right)^2 - 2 \Omega^2 \omega_E^2   \;\;\;,
\end{equation}

\begin{equation}\label{anis56}
H_1 = - \Omega^2 \omega_E^2   \;\;\;,
\end{equation}

\noindent
and

\begin{equation}\label{anis57}
H_2 = \Omega^2 \omega_E^2   \;\;\;,
\end{equation}

\noindent
Rearranging Eq. (\ref{anis54}) we have

\begin{equation}\label{anis58}
H_0 \left(H_0 - H_1 + H_2\right) \left\{H_0 \left(H_0 - H_1 + H_2\right) + 3 \Omega^4 \omega_E^4\right\} = 0   \nonumber   \;\;\;.
\end{equation}

\noindent
It implies the following resonance conditions

\begin{equation}\label{anis59}
\omega_m = \omega   \;\;\;,
\end{equation}

\begin{equation}\label{anis60}
\omega_L = \sqrt{\omega^2 + \frac{\Omega^2}{2}} - \sqrt{\frac{\Omega^2}{2}}   \;\;\;
\end{equation}

\noindent
and

\begin{equation}\label{anis61}
\omega_R = \sqrt{\omega^2 + \frac{\Omega^2}{2}} + \sqrt{\frac{\Omega^2}{2}}   \;\;\;
\end{equation}

\begin{figure}[!htb]
\includegraphics[width=16cm,angle=0,clip]{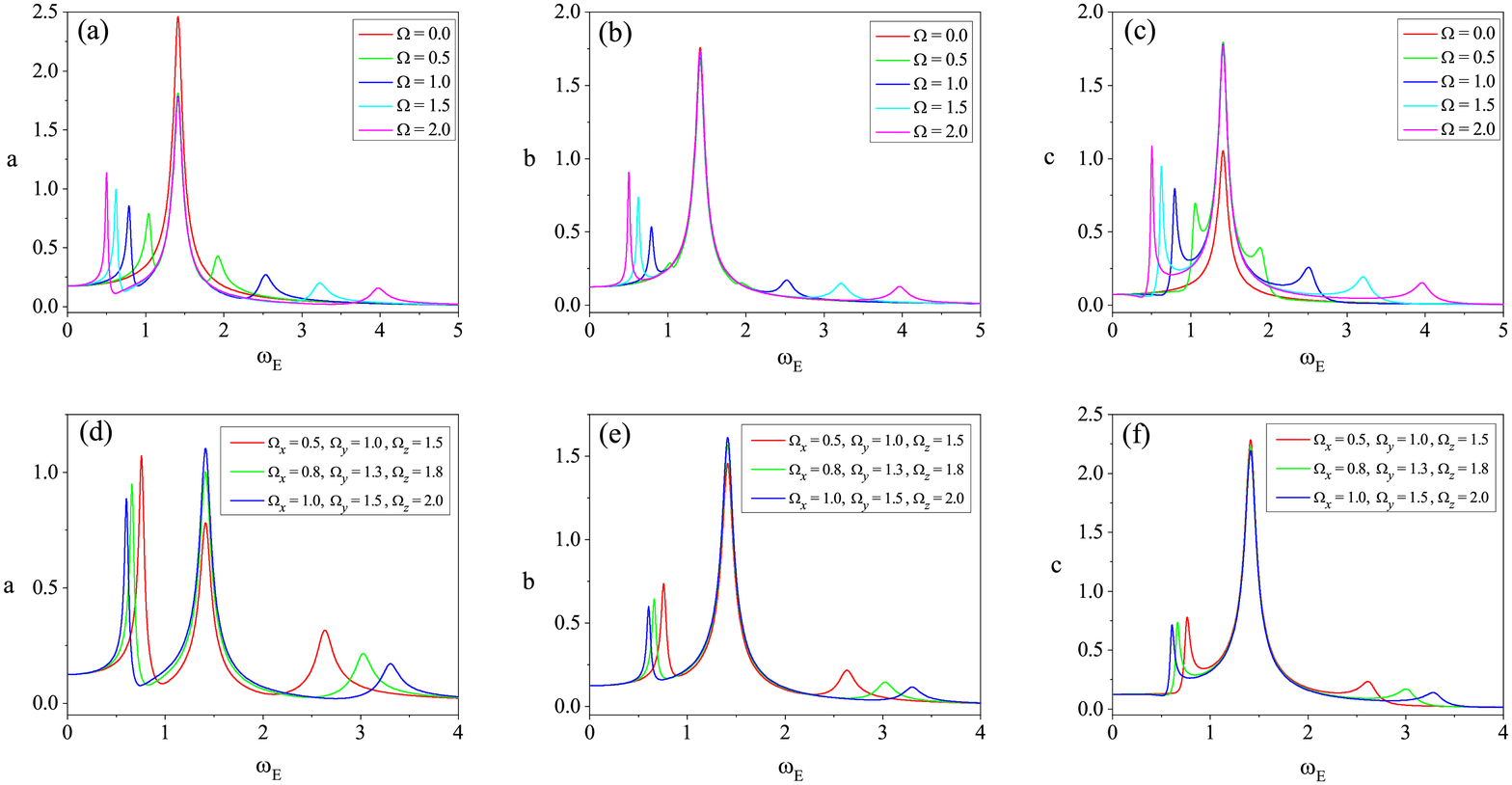}
\caption{{\footnotesize Plot of $a$, $b$ and $c$ vs. $\omega_E$ for the parameter set: $\omega^2 = 2.0, \gamma = 0.1$, $E_{0x} = E_{0y} = E_{0z} = 0.25$ and $\phi_x = \phi_y = \phi_z = 0.0$ (Units are arbitrary).}}
\label{fig.8}
\end{figure}

\noindent
These conditions are consistent with panels (a), (b) and (c) in Fig.\ref{fig.8}. We compare the above conditions with the exact results in Table \ref{tab.4}. It shows a fair agreement between them.

\begin{table}[ht]
\caption{Comparison between theoretically calculated resonating frequency and the exact result
for the driven damped isotropic harmonic oscillator}
\begin{center}
\begin{tabular}{|c|c|c|c|c|c|c|}
\hline
Value of &
\multicolumn{2}{|c|}{Resonance at $\omega_L$} &
\multicolumn{2}{|c|}{Resonance at $\omega_m$} &
\multicolumn{2}{|c|}{Resonance at $\omega_R$} \\
\cline{2-7}
$\Omega$ & Theoretical & Exact & Theoretical & Exact & Theoretical & Exact \\ 
\hline
0.5 & 1.104 & 1.039 & 1.412 & 1.409 & 1.811 & 1.919 \\
\hline
1.0 & 0.874 & 0.789 & 1.412 & 1.409 & 2.288 & 2.529 \\
\hline
1.5 & 0.707 & 0.620 & 1.412 & 1.409 & 2.828 & 3.230 \\
\hline
2.0 & 0.586 & 0.500 & 1.412 & 1.409 & 3.414 & 3.970 \\
\hline
\end{tabular}
\end{center}
\label{tab.4}
\end{table}

\noindent
Another point is to be noted here. It is apparent in the resonance conditions that in the presence of a magnetic field, a vibrational motion may be composed of three frequencies and one ($\omega_m$) of them may not depend on the applied field as a signature of the velocity dependent coupling in the equation of motion. This is also true even for the case, $\omega_x = \omega_y = \omega_z = \omega$, $\Omega_x \neq \Omega_y \neq \Omega_z \neq \Omega$ as shown in panels (d), (e) and (f) of Fig.\ref{fig.8}. Thus the appearance of three peaks for the isotropic harmonic oscillator has a similarity with the Normal Zeeman Effect in the sense that the middle peak does not depend on the applied magnetic field. One may find all the peaks making use of oscillating electric field which may be unpolarized or polarized at any of the three directions. However, for the driven damped isotropic harmonic oscillator in the presence of a magnetic field along the $z$-direction, Eqs. (\ref{anis60}-\ref{anis61}) reduce to Eqs. (\ref{eqr31}-\ref{eqr32}) and one may find three peaks at the frequencies given by Eqs. (\ref{anis59}, (\ref{eqr31}-\ref{eqr32}). This is also similar to the Normal Zeeman Effect. To avoid any confusion we would mention here that if the driven damped three dimensional anisotropic oscillator experiences a magnetic field along the $z$-direction then the characteristics of the spectrum also may mimic the Normal Zeeman Effect. Shortly we will show that one may determine approximately the field dependent resonating frequencies. For these cases the motion along the $z$-direction is decoupled from the motion in $x$-$y$ plane. Then three peaks may appear for the polarized electric field which may lying either in $x$-$z$ or $y$-$z$ plane. Of course one may find all the three peaks making use of unpolarized electric field. Finally, it is to be noted here that the position of the three peaks may depend on the direction of the applied field (if it is not along the $z$-direction) for the three dimensional anisotropic oscillator as implied in Fig.\ref{fig.7}.

\begin{figure}[!htb]
\includegraphics[width=16cm,angle=0,clip]{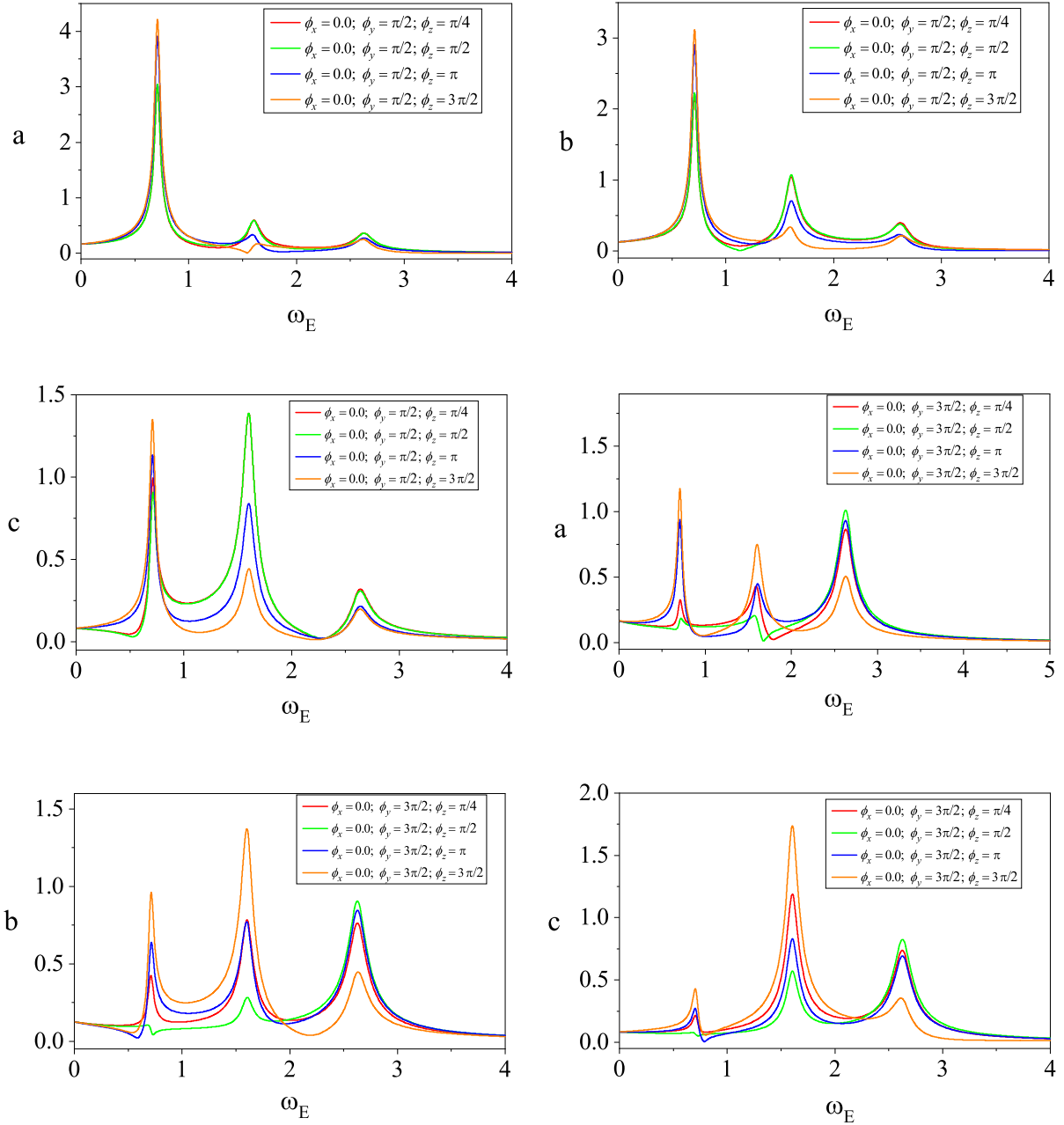}
\caption{{\footnotesize Plot of $a$, $b$ and $c$ vs. $\omega_E$ for different values of $\phi_x, \phi_y$ and $\phi_z$ along with the parameter set: $\omega_x^2 = 1.5, \omega_y^2 = 2.0, \omega_z^2 = 3.0, \gamma = 0.1$ and $E_{0x} = E_{0y} = E_{0z} = 0.25$ (Units are arbitrary).}}
\label{fig.9}
\end{figure}

We are now in a position to demonstrate the effect of phase difference among the components of the input signal on the output signal. Fig.\ref{fig.9} has been included in this context. Again this figure implies that the phase difference between the input signals may modulate amplitude of the output signals  as a signature of the interference between the input signals through the velocity dependent coupling. As a result of that the special features for the two dimensional case is also continued for three dimensional motion. One more additional feature, anti resonance may appear at nearby the resonating frequency for the latter one.

\noindent
Before leaving this section we show that one may determine the resonance condition for the driven damped anisotropic oscillator at relatively simple situation such as $\omega_z = 0$,  $\Omega_x = \Omega_y = 0$ and $\Omega_z =\Omega $. Then the relevant amplitudes and phases of the output signal can be read from Eqs. (\ref{anis8}, \ref{anis9}, \ref{anis11}, \ref{anis12}) as

\begin{equation}\label{df3}
a = \frac{\sqrt{H_1^2 + H_2^2}}{H_0}   \;\;\;,
\end{equation}

\begin{equation}\label{df4}
b = \frac{\sqrt{H_3^2 + H_4^2}}{H_0}   \;\;\;,
\end{equation}

\begin{equation}\label{df5}
\tan \phi_1 = \frac{H_2}{H_1}   \;\;\;,
\end{equation}

\noindent
and

\begin{equation}\label{df6}
\tan \phi_2 = \frac{H_4}{H_3}   \;\;\;.
\end{equation}

\noindent
Here we have used

\begin{equation}\label{df7}
H_0 = H_{0x} H_{0y} + \gamma^2 \Omega^2 \omega_E^4 \left(\omega_x^2 + \omega_y^2 - 2 \omega_E^2\right)^2   \;\;\;,
\end{equation}

\begin{equation}\label{df8}
H_1 = \frac{q}{m} \left(\omega_x^2 - \omega_E^2\right) \left(\omega_y^2 - \omega_E^2\right) \left\{E_{0x} H_{0y} + E_{0y} \gamma \Omega \omega_E^2 \left(\omega_x^2 + \omega_y^2 - 2 \omega_E^2\right)\right\}   \;\;\;,
\end{equation}

\begin{equation}\label{df9}
H_2 = \frac{q}{m} \omega_E \left(\omega_y^2 - \omega_E^2\right) \left\{E_{0x} H_{0y} \gamma - E_{0y} H_{0x} \Omega + \gamma \Omega \omega_E^2 \left(\omega_x^2 + \omega_y^2 - 2 \omega_E^2\right) \left(E_{0y} \gamma + E_{0x} \Omega\right)\right\}   \;\;\;,
\end{equation}

\begin{equation}\label{df10}
H_3 = \frac{q}{m} \left(\omega_x^2 - \omega_E^2\right) \left(\omega_y^2 - \omega_E^2\right) \left\{E_{0y} H_{0x} - E_{0x}\gamma \Omega \omega_E^2 \left(\omega_x^2 + \omega_y^2 - 2 \omega_E^2\right)\right\}   \;\;\;,
\end{equation}

\begin{equation}\label{df11}
H_4 = \frac{q}{m} \omega_E \left(\omega_x^2 - \omega_E^2\right) \left\{E_{0y} H_{0x} \gamma + E_{0x} H_{0y} \Omega + \gamma \Omega \omega_E^2 \left(\omega_x^2 + \omega_y^2 - 2 \omega_E^2\right) \left(E_{0y} \Omega - E_{0x} \gamma\right)\right\}   \;\;\;.
\end{equation}

\noindent
with

\begin{equation}\label{df12}
H_{0x} = \left(\omega_x^2 - \omega_E^2\right)^2 \left(\omega_y^2 - \omega_E^2\right) + \gamma^2 \omega_E^2 \left(\omega_y^2 - \omega_E^2\right) - \Omega^2 \omega_E^2 \left(\omega_x^2 - \omega_E^2\right)   \;\;\;,
\end{equation}

\noindent
and

\begin{equation}\label{df13}
H_{0y} = \left(\omega_y^2 - \omega_E^2\right)^2 \left(\omega_x^2 - \omega_E^2\right) + \gamma^2 \omega_E^2 \left(\omega_x^2 - \omega_E^2\right) - \Omega^2 \omega_E^2  \left(\omega_y^2 - \omega_E^2\right)  \;\;\;,
\end{equation}

\noindent
It is to be noted here that for the purpose of the determination of the resonance condition we have chosen no phase difference between the driving components. However, one can show that for $\omega_x = \omega_y$, Eqs. (\ref{df3}-\ref{df6}) reduce to Eqs. (\ref{eqr11}-\ref{eqr14}). It is an important check for this section.

\begin{figure}[!htb]
\includegraphics[width=16cm,angle=0,clip]{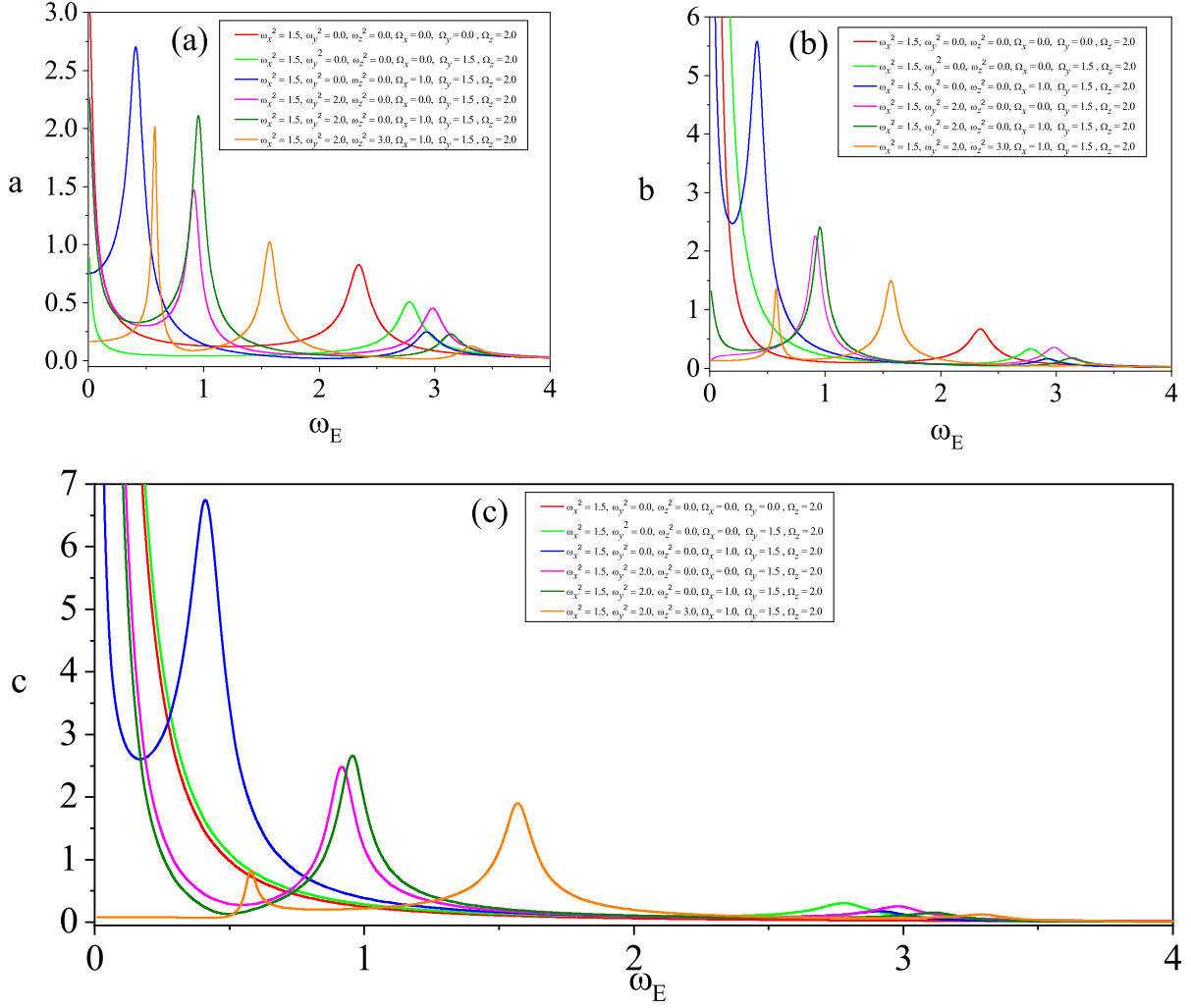}
\caption{{\footnotesize Plot of $a$, $b$ and $c$ vs. $\omega_E$ for the parameter set: $\gamma = 0.1$, $E_{0x} = E_{0y} = E_{0z} = 0.25$ and $\phi_x = \phi_y = \phi_z = 0.0$ (Units are arbitrary).}}
\label{fig.10}
\end{figure}

Following the earlier calculation we now find the resonance condition. Since the resonance phenomenon may appear at $\gamma \rightarrow 0$ then the amplitude of the output signal may be maximum around the following condition

\begin{equation}\label{df12a}
\left(\omega_x^2 - \omega_E^2\right) \left(\omega_y^2 - \omega_E^2\right) - \Omega^2 \omega_E^2 = 0   \;\;\;.
\end{equation}

\noindent
If $\omega_y = 0$ then the solution of the above can be read as

\begin{equation}\label{df12b0}
\omega_E = \sqrt{\omega_x^2 + \Omega^2}   \;\;\;.
\end{equation}

\noindent
and

\begin{equation}\label{df12b}
\omega_E = 0   \;\;\;.
\end{equation}

\noindent
Similarly for $\omega_x = 0$ we have

\begin{equation}\label{df12c0}
\omega_E = \sqrt{\omega_y^2 + \Omega^2}   \;\;\;.
\end{equation}

\noindent
and
\begin{equation}\label{df12c}
\omega_E = 0   \;\;\;.
\end{equation}

\noindent
The above Eqs. (\ref{df12b0}-\ref{df12c}) imply that for the one dimensional harmonic oscillator in the presence of a magnetic field (which is perpendicular to the direction of the oscillator) only one resonance peak may appear as shown red curve in panel (a) of Fig.10. For the given parameter set for this curve, the location of the peak at $\omega_E = 2.345$. This is very much consistent with the theoretical value. To avoid any confusion we would mention here that the  peak at $\omega_E = 0$
as implied by Eq.(\ref{df12b}) is corresponding to the diverging motion in the presence of the constant electric field. One may verify it easily considering the relevant equations of motion. However,
Fig.10 is a typical demonstration of steady state dynamics of  driven damped anisotropic oscillators. It shows that if the field is tilted from the perpendicular direction then one more additional peak may appear. Panels (b) and (c) imply that the additional peak appears as a signature of a complex motion whose projection in $y-z$ plane is like a damped driven cyclotron motion. Then continuation of the discussion for a higher dimensional oscillator is straight forward. 

Then for $\omega_x = \omega_y = 0$ one may show easily from Eq. (\ref{df12a}) that

\begin{equation}\label{df12d}
\omega_E = \Omega^2   \;\;\;.
\end{equation}

\noindent
Thus it constitutes an important check for the present approximation calculation. We now consider the condition when both $\omega_x$ and $\omega_y$ are not zero. If $\omega_y \ll \omega_E$ then one may show easily from Eq. (\ref{df12a}) that

\begin{equation}\label{df18}
\omega_R \simeq \sqrt{\omega_x^2 + \Omega^2}   \;\;\;.
\end{equation}

\noindent
Similarly for $\omega_x \ll \omega_E$ we get

\begin{equation}\label{df19}
\omega_R \simeq \sqrt{\omega_y^2 + \Omega^2}   \;\;\;.
\end{equation}

\begin{table}[ht]
\caption{Comparison between theoretically calculated resonating frequency and the exact result for the driven damped an-isotropic harmonic oscillator. For the relation, $\omega_R \simeq \sqrt{\omega_x^2 + \Omega^2}$, $\omega_x^2 = 2.0$ and $\omega_y^2 = 0.1$. Similarly for $\omega_R \simeq \sqrt{\omega_y^2 + \Omega^2}$, $\omega_x^2 = 0.1$ and $\omega_y^2 = 3.0$.}
\begin{center}
\begin{tabular}{|c|c|c|c|c|}
\hline
Value &
\multicolumn{2}{|c|}{Resonance at $\omega_R \simeq \sqrt{\omega_x^2 + \Omega^2}$} &
\multicolumn{2}{|c|}{Resonance at $\omega_R \simeq \sqrt{\omega_y^2 + \Omega^2}$} \\
\cline{2-5}
of $\Omega$ & Theoretical & Exact & Theoretical & Exact \\ 
\hline
0.5 & 1.500 & 1.500 & 1.803 & 1.799 \\
\hline
1.0 & 1.732 & 1.740 & 2.000 & 2.000 \\
\hline
1.5 & 2.061 & 2.069 & 2.291 & 2.299 \\
\hline
2.0 & 2.449 & 2.460 & 2.646 & 2.649 \\
\hline
\end{tabular}
\end{center}
\label{tab.5}
\end{table}
\begin{figure}[!htb]
\includegraphics[width=16cm,angle=0,clip]{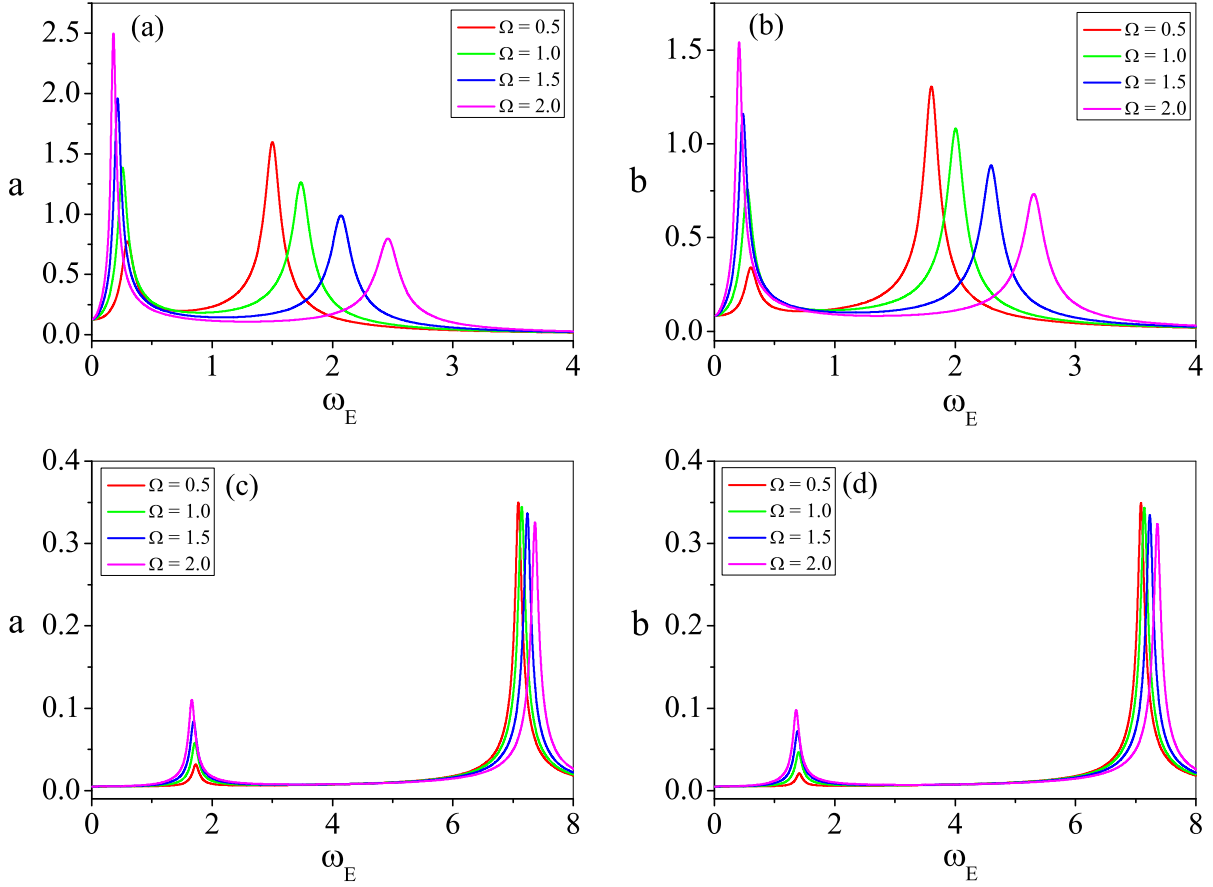}
\caption{{\footnotesize Plot of $a$ and $b$ vs. $\omega_E$ for different values of $\Omega$ at various limits of $\omega_x^2$ and $\omega_y^2$: (a) $\omega_x^2 = 2.0, \omega_y^2 = 0.1$, (b) $\omega_x^2 = 0.1, \omega_y^2 = 3.0$, (c) $\omega_x^2 = 50.0, \omega_y^2 = 3.0$, and (d) $\omega_x^2 = 2.0, \omega_y^2 = 50.0$. Other common parameters are: $E_{0x} = E_{0y} = 0.25$ (Units are arbitrary).}}
\label{fig.11}
\end{figure}

\begin{table}[ht]
\caption{Comparison between theoretically calculated resonating frequency and the exact result for the driven damped an-isotropic harmonic oscillator. For the relation, $\omega_L \simeq \frac{\omega_x \omega_y}{\sqrt{\omega_x^2 + \Omega^2}}$, $\omega_x^2 = 50.0$ and $\omega_y^2 = 3.0$. Similarly for  $\omega_L \simeq \frac{\omega_x \omega_y}{\sqrt{\omega_y^2 + \Omega^2}}$,  $\omega_x^2 = 2.0$ and $\omega_y^2 = 50.0$.}
\begin{center}
\begin{tabular}{|c|c|c|c|c|}
\hline
Value &
\multicolumn{4}{|c|}{Corresponding to the left peak} \\
\cline{2-5}
of &
\multicolumn{2}{|c|}{Resonance at $\omega_L \simeq \frac{\omega_x \omega_y}{\sqrt{\omega_x^2 + \Omega^2}}$} &
\multicolumn{2}{|c|}{Resonance at $\omega_L \simeq \frac{\omega_x \omega_y}{\sqrt{\omega_y^2 + \Omega^2}}$} \\
\cline{2-5}
$\Omega$ & Theoretical & Exact & Theoretical & Exact \\ 
\hline
0.5 & 1.723 & 1.730 & 1.407 & 1.409 \\
\hline
1.0 & 1.715 & 1.709 & 1.400 & 1.399 \\
\hline
1.5 & 1.707 & 1.689 & 1.393 & 1.379 \\
\hline
2.0 & 1.698 & 1.659 & 1.387 & 1.360 \\
\hline
\end{tabular}
\end{center}
\label{tab.6}
\end{table}

\noindent
Again for $\omega_x \gg \omega_E$, we have from Eq. (\ref{df12a})

\begin{equation}\label{df20}
\omega_L \simeq \frac{\omega_x \omega_y}{\sqrt{\omega_x^2 + \Omega^2}}   \;\;\;.
\end{equation}

\noindent
Similarly for $\omega_y \gg \omega_E$, we get

\begin{equation}\label{df21}
\omega_L \simeq \frac{\omega_x \omega_y}{\sqrt{\omega_y^2 + \Omega^2}}   \;\;\;.
\end{equation}

\noindent
To check the accuracy of our calculation we have demonstrated the exact results (\ref{df3}-\ref{df6}) in Fig.\ref{fig.11}. The resonance conditions according to this figure are compared with the analytical results in Table \ref{tab.5} and \ref{tab.6}. These show that there is a very good agreement between the approximate and the exact results. Further more, one may explain any distinguishable feature such as the peak height of asymmetric spectrum of amplitudes considering the interference between the input signal.

\subsection{The energetics:}

Extension of Eqs. (\ref{en2}-\ref{en4}) for three dimensional harmonic oscillator can be read as

\begin{equation}\label{eneq}
\langle P \rangle = \frac{1}{2} m \gamma \omega_E^2 \left(a^2 + b^2 + c^2\right)   \nonumber   \;\;\;,
\end{equation}

\begin{equation}\label{eneq1}
\langle E \rangle = \frac{1}{2} m \left(\omega_x^2 + \omega_E^2\right) \frac{1}{2} a^2 + \frac{1}{2} m \left(\omega_y^2 + \omega_E^2\right) \frac{1}{2} b^2 +  \frac{1}{2} m \left(\omega_z^2 + \omega_E^2\right) \frac{1}{2} c^2   \nonumber   \;\;\;
\end{equation}

\noindent
and

\begin{equation}\label{eneq2}
Q = \frac{\left(\omega_x^2 + \omega_E^2\right) a^2 + \left(\omega_y^2 + \omega_E^2\right) b^2 + \left(\omega_z^2 + \omega_E^2\right) c^2}{2 \gamma \omega_E \left(a^2 + b^2 + c^2\right)}   \nonumber   \;\;\;
\end{equation}

\noindent
Making use of Eqs. (\ref{anis8}-\ref{anis10}) into the above relations one may find the role of magnetic field in the process of energy storing.

\section{Conclusion}
In the present study we have considered various aspects of driven damped harmonic oscillator in the presence of a magnetic field. Our investigation includes the following points.

(A) Dynamics of a two dimensional harmonic oscillator (having same frequency along both the directions) in the presence of magnetic and dissipative forces:

(i) The exact solution of the equations of motion in the absence of damping exhibits a time dependent position which is composed of two frequencies as a signature of the cross effect of the non conservative magnetic force. Then we have determined the condition for a simple periodic motion.

(ii) We have determined the solutions of equations of motion at weak damping limit which  exhibits that the damped oscillation is a composed of two frequencies. These solutions reduce to all the standard results  for the cases like damped harmonics oscillator and cyclotron motion, respectively. 

Thus the above cases may imply the resonance conditions for the periodically driven damped harmonic oscillator in the presence of a magnetic field. Regarding this the following major points are given below.

(B) Dynamics at steady state for the periodically driven damped two dimensional harmonic oscillator in the presence of a magnetic field:

(i) We have determined the relevant amplitudes and phase constants for the oscillation at steady state.  The values of the amplitudes at resonance condition become infinite in the absence of damping. It is to be noted here that  for this case, the phase shift between the input and output signals at the resonance condition is similar as that of the driven damped harmonic oscillator. It proofs indirectly that the finite value of the amplitude at the resonance condition in the presence of damping is solely due to the dissipation of energy. In other words, the phase shift has no significant role in this context. 

(ii) Magnetic field induces an asymmetric splitting of the spectrum of output signal with two peaks.

(iii) The relevant resonance conditions has been determined with a very good approximation which describes how the resonating frequency depends on the magnetic field and the force constant of the harmonic oscillator. This leads to have the resonance condition for the driven damped cyclotron motion.

(iv) There is a magnetic field induced phase shift between the input and output signals bears a clear signature of the magnetic field induced breakdown of the equivalence of the two dimensional motion. 

(v) Using the phase difference between the components of the driving field one may modulate the amplitude of the oscillation at the steady state. As a consequence of the interference between the two driving components through the velocity dependent coupling the following special features may appear.  (a) The peak height at higher frequency may be higher compared to the other (b) One of the two pics may disappear for a certain phase difference. (c) The anti resonance phenomenon may occur. Even it may appear for a driven damped cyclotron motion where the system with purely non conservative force fields is driven by an electric field. 
It is to be noted here that for these cases both resonance and anti resonance phenomena may appear at the same driving frequency depending upon the phase difference between the input signals. But for a three dimensional anisotropic oscillator the anti resonance may appear at near by the resonating frequency.
 
(vi) Our calculation shows that the stored average energy closely mimics the variation of amplitude as a function of driving frequency but the average power does not follow this pattern. At the resonance condition, the latter decreases monotonically with increase in the strength of the applied field but the former pass through a minimum. For a given damped harmonic oscillator the efficiency like quantity in the energy storing process does not depend on the parameters related to driving electric field. But it depends on the damping strength and the frequency of the oscillator, respectively.

(C) Dynamics at steady state for the periodically driven damped three dimensional harmonic oscillator in the presence of a magnetic field at arbitrary direction:

(i) For this very general case, we have calculated the relevant amplitudes and the phase constants for the steady state dynamics. Here we find that magnetic field may induce one more additional peak compared to the previous case. It is interesting to be noted here that the position of the additional peak for the isotropic harmonic oscillator does not depend on the strength and the direction of the applied magnetic field . Thus this observation has a similarity with the Normal Zeeman Effect. Again it is to be noted here that if a driven damped isotropic harmonic oscillator experiences a magnetic field along $z$-direction then one may find three peaks and the phenomenon is also similar to the Normal Zeeman Effect. To avoid any confusion we would mention here that if the driven damped three dimensional an-isotropic oscillator experiences a magnetic field along the $z$-direction then the characteristics of the spectrum also may mimic the Normal Zeman Effect.  Finally, it is to be noted here that the position of the  three peaks may depend on the direction of the applied field (if it is not along the $z$-direction) for the three dimensional an-isotropic oscillator 

(ii) The generalization of the problem restricts us to determine the relevant resonance conditions only for the special cases like (1) the periodically driven damped isotropic harmonic oscillator in the presence of a magnetic field whose all the components are same, (2) the periodically driven damped two dimensional harmonic oscillator (having different force constant along each direction) whose motion is confined in $x-y$ plane in the presence of a magnetic field along $z$-direction and (3) the periodically driven damped one dimensional harmonic oscillator in the presence of a magnetic field which is perpendicular to the direction of the harmonic motion.

(iii) Finally, we have determine how the quantities like the phase shift, the average power and the average stored energy depend on the direction of the magnetic field, the damping strength and the frequencies of the three dimensional oscillator.

Before leaving this section we would mention the possible applications of the present study. In this context one may address the following issues. First, effect of magnetic field on the vibrational resonance.
Second, the magnetic field may induce a phase shift between the input and output signals even in the absence of damping. Thus investigation on the modulation of the refractive index of a material by virtue of the Lorentz force may be an worthy issue. Third, modulation of the frequency as well as energy of a harmonic oscillator by an applied magnetic field clearly requires a detail study on the thermally activated barrier crossing dynamics in the presence of an electromagnetic field. One may be interested to check whether the asymmetric splitting of the spectrum of the barrier crossing rate constant occurs or not in the presence of the field. Finally, in a recent study \cite{shmpre} we have shown that a colored noise can recognize the character of a dynamical system in terms of autonomous stochastic resonance which caries the sense of the dynamical resonance. Thus investigation on the effect of magnetic field on the autonomous stochastic resonance may be an worthy issue. These issues are in progress and shortly may appear elsewhere.

\end{document}